\documentclass[prd,twocolumn,showpacs,preprintnumbers,twoside]{revtex4}
\usepackage{amsmath}
\usepackage{amsfonts}
\usepackage{amssymb}
\usepackage{bm}
\usepackage{graphicx}  

\begin{document}

\bibliographystyle{/home/tkachev/TeX/local/h-physrev3}

\preprint{}
\title{Turbulent  Thermalization}
\author{Raphael Micha}
\email{micha@itp.phys.ethz.ch}
\affiliation{%
Theoretische Physik, 
    ETH Z\"urich, 
    CH-8093 Z\"urich, 
    Switzerland 
}%
\author {Igor I. Tkachev}
\email{igor.tkachev@cern.ch}
\affiliation
    {%
      Department of Physics, Theory Division, CERN, CH-1211 Geneva 23, 
      Switzerland \\
      and \\
      Institute for Nuclear Research of the Russian Academy of
      Sciences, 117312, Moscow, Russia
    }%


\begin{abstract}
  We study, analytically and with lattice simulations, the decay of coherent
  field oscillations and the subsequent thermalization of the resulting
  stochastic classical wave-field. The problem of reheating of the Universe
  after inflation constitutes our prime motivation and application of the
  results.  We identify three different stages of these processes. During the
  initial stage of ``parametric resonance'', only a small fraction of the
  initial inflaton energy is transferred to fluctuations in the physically
  relevant case of sufficiently large couplings.  A major fraction is
  transfered in the prompt regime of driven turbulence.  The subsequent long
  stage of thermalization classifies as free turbulence.  During the turbulent
  stages, the evolution of particle distribution functions is self-similar. We
  show that wave kinetic theory successfully describes the late stages of our
  lattice calculation.  Our analytical results are general and give estimates
  of reheating time and temperature in terms of coupling constants and initial
  inflaton amplitude.
\end{abstract}


\maketitle

\section{Introduction}

Field theoretical systems which are a long way from thermal equilibrium have
been studied intensely in recent years. The particular problem of how and when
such systems approach equilibrium stretches beyond obvious fundamental interest
and finds many practical applications. In high-energy physics understanding of
these processes is crucial for applications to heavy ion collisions and to
cosmology of the early universe. The first topic gains further importance in
light of the current and future experimental search for a quark-gluon-plasma
at RHIC and at the forthcoming LHC. The second application, our main
interest in this paper, is related to the problem of reheating of the
universe after cosmological inflation.

Inflation provides a solution to the flatness and the horizon problems of
standard cosmology \cite{lindebook,kolbbook,lythbook} and explains the
generation of initial density perturbations - the seeds of galaxies and
large-scale structure in our universe. During inflation the universe is in a
vacuum-like state. At the end of inflation all energy density is stored in a
Bose condensate, the coherently oscillating "inflaton" field. This state is
highly unstable: parametric, tachyonic or strong non-adiabatic particle
creation triggers a fast and explosive decay of the inflaton. This process,
dubbed preheating \cite{Kofman:1994rk,Shtanov:1995ce}, is currently well
understood
\cite{Khlebnikov:1996mc,Khlebnikov:1997wr,Khlebnikov:1997zt,Prokopec:1997rr,
Kofman:1997yn,Garcia-Bellido:1998wm, Felder:2000hj,Giudice:1999fb}. A generic
feature is a strong and fast amplification of fluctuation fields at low
momenta, which may lead to various interesting physical effects during and
after preheating. These include non-thermal phase transitions
\cite{Kofman:1995fi,Tkachev:1996md, Riotto:1996vi,Khlebnikov:1998sz} with
possible formation of topological defects
\cite{Tkachev:1998dc,Kasuya:1997ha,Rajantie:2000fd,Ray:2001cd,Copeland:2002ku},
creation of super-heavy particles \cite{Felder:1998vq,Giudice:1999fb},
generation of high-frequency gravitational waves \cite{Khlebnikov:1997di},
etc.

The explosive stage of inflaton decay ends when the rate of interactions of
created fluctuations among themselves and with the inflaton becomes comparable
to the inflaton decay rate
\cite{Khlebnikov:1996mc,Khlebnikov:1997wr,Khlebnikov:1997zt,Prokopec:1997rr}.
The understanding of the subsequent stages of relaxation towards equilibrium,
of the thermalization processes and the calculation of the final equilibrium
temperature is important for various applications as it links the inflationary
phase with that of standard cosmology. Among those one can list baryogenesis
\cite{Kolb:1996jt,Kolb:1998he,Giudice:1999fb,Garcia-Bellido:1999sv,
Garcia-Bellido:2001cb} and the problem of over-abundant gravitino production
in supergravity models
\cite{Ellis:1982yb,Ellis:1985er,Kallosh:1999jj,Giudice:1999yt,Giudice:1999am,
Maroto:1999ch}.  It determines the abundances of other relics, like
super-heavy dark matter
\cite{Chung:1998zb,Kuzmin:1998uv,Kuzmin:1998kk,Chung:1999ve}, or axino dark
matter \cite{Covi:2004rb}.

Knowledge of the reheating temperature is also important for fixing
constraints on the inflationary model from Cosmic Microwave Background (CMBR)
anisotropy \cite{Liddle:1993fq,Dodelson:2003vq,Liddle:2003as,Feng:2003nt}.  In
some models cosmologically important curvature perturbations may be even
generated during the process of thermalization
\cite{Hamazaki:1996ir,Dvali:2003em,Kofman:2003nx,Enqvist:2003uk,
Matarrese:2003tk}. 
Last but not least: the reheating temperature should be larger
than about one GeV to ensure that the standard Big Bang Nucleosynthesis
\cite{kolbbook,Olive:1999ij} is not hampered.

There have been many efforts and successes in the understanding of the
non-equilibrium dynamics and relaxation of field theories, see e.g.
Refs. \cite{Son:1996uv,Semikoz:1996ty,Aarts:2000mg,Davidson:2000er,
Salle:2000hd,Calzetta:2000kn,Borsanyi:2000ua,Felder:2000hr,Borsanyi:2000pm,
Bodeker:2000pa,Aarts:2002dj,Berges:2002cz,Borsanyi:2003ib,Boyanovsky:2003tc,
Baacke:2003bt,Ikeda:2004in}.  However, the leading asymptotic dynamics towards
equilibrium remained rather less understood and developed.

The main problem for the theoretical understanding of reheating is that
initially the occupation numbers are very large, of order of the inverse
coupling constant. In addition, in many inflationary models the zero mode does
not decay completely during preheating. Therefore, a simple perturbative
approach is not justified.  On the other hand, in this regime, a description in
terms of classical field theory is valid \cite{Khlebnikov:1996mc}, and the
whole process (including preheating), can be studied by classical lattice
simulations.

Recently, we employed this method to show \cite{Micha:2002ey,Micha:2003ws}
that the classical reheating of a massless $\Phi^4$-theory in 3+1 dimensions
is characterized by a \textit{ turbulent} and \textit{ self-similar} evolution
of distribution functions towards equilibrium. The shape of the spectra, as
well as the self-similar dynamics, could be understood within the framework of
wave kinetic theory.  This made it possible to estimate reheating time and
temperature, which turned out to coincide parametrically with the results of
the simple perturbative approach.

Turbulence appears in a large variety of non-equilibrium-phenomena in nature
(see Refs.~\cite{ZakBook,lvovbook,FrischBook} for a general introduction). It
was first discussed for fluids, in the regime of large Reynolds numbers
(velocities), where viscosity is subdominant.  Kolmogorov identified
turbulence in this regime \cite{K41:1,K41:2} as a statistically scale
invariant flow of spectral energy mediated by vortex interactions.  The same
dynamical structure may appear in systems of coupled waves, e.g. on fluid
surfaces or for coupled fields in a plasma \cite{Z67,ZakBook,lvovbook}. In
this case the cascade is mediated by wave interactions and the phenomenon has
been called \textit{wave turbulence}.

If there exists an active (stationary) source of energy in momentum space, the
turbulence is called \textit{driven} (stationary).  When the source is
switched off after the stage of activity, the freely propagating energy
cascade is often referred to as \textit{free} turbulence. If the kinetic
description is applicable, the energy cascade is called \textit{weak
turbulence}. Otherwise one is facing a \textit{strong turbulence}.

One may expect that the concept of turbulence should be relevant for the
problem of reheating \cite{Khlebnikov:1996mc,Son:1996uv} already on general
grounds. Indeed, the source of energy, localized in the "infra-red" is present
initially.  It is represented by the inflaton field in the problem at
hands. To complete the argument, we note that as the final outcome of the
evolution one should expect cascading of energy towards a significantly
separated region of ``ultra-violet'', high momentum modes.

The goal of the current paper is twofold. First, we want to apply the wave
kinetic theory of turbulence to the problem of Universe reheating after
inflation. We derive general formulas for the spectra of turbulent
distributions and for the self-similar evolution towards equilibrium. This
enables us to give asymptotic estimates of reheating time and temperature in
Minkowski space as well as in Friedmann universe.  

Second, we want to test and confront these ideas to numerical lattice
calculations. For our numerical integrations we have chosen the simplest
``chaotic'' inflationary model \cite{Linde:1983gd}. While the initial
``preheating'' stage in other inflationary models, e.g. in hybrid inflation
\cite{Linde:1994cn} may exhibit important differences
\cite{Garcia-Bellido:1998wm,Felder:2000hj,Micha:1999wv} with this model, we
expect the subsequent turbulent stages to be more universal.
 
 We start lattice integration from ``vacuum'' initial conditions for
fluctuations in a background of classical oscillating inflaton field.  We
observe the initial parametric resonance stage when the energy in fluctuations
is growing exponentially with time. This stage terminates when re-scattering
of waves out of the resonance band becomes important
\cite{Khlebnikov:1996mc,Khlebnikov:1997zt}.  In the physically relevant case
of sufficiently large couplings this happens rather early, when only a small
fraction of initial inflaton energy is transferred to fluctuations
\cite{Khlebnikov:1997zt,Prokopec:1997rr,Kofman:1997yn}. At this point a state
of stationary turbulence should be established that is driven by the
zero-mode. On general grounds, it can be deduced that during this stage the
energy in fluctuations should grow linearly with time. This behavior is
confirmed by the results of our numerical simulations.  The stage of
stationary turbulence should terminate when the energy left out in the
zero-mode becomes smaller than the energy stored in created "particles".  From
this moment of time, the transport of energy from the source is negligible and
we observe free turbulence with self-similar evolution of particle
distributions towards thermal equilibrium.

The first stage of driven turbulence is prompt and gives the main mechanism by
which energy is drawn out of the zero-mode, e.g. out of the inflaton field.
The identification of this constitutes one of the new results of the present
paper, as opposed to the common opinion that the main mechanism is a
"parametric resonance''. The second stage of free turbulence is very long
and can be analytically described as self-similar evolution. This is another
new result and diffuses some existing claims and hopes that "parametric
resonance" may bring a system to thermal equilibrium on a very short time
scale.

Overall, the kinetic description and the results of lattice simulations are in
rather good agreement with each other.  This indicates that the regime of weak
wave turbulence may be already achieved on the lattice.

The paper is organized as follows. In Section \ref{NumSim:One-Field} we review
the results of our numerical simulation of reheating in the simplest $\lambda
\Phi^4$ model to get familiar with concepts, problems and the typical
dynamical behavior of the systems of interest. In Sections
\ref{Turbulentreheating}, \ref{sec4} we apply the theory of wave turbulence to
the problem of reheating in general.  In Section \ref{TwoFields} we present
our numerical simulations.  In Section \ref{CompLatt2Kin} we compare lattice
results with the kinetic approach and discuss the applicability of the
latter. In Section \ref{PhysApplications} we discuss some physical
applications of our results, in particular the thermalization in the
self-similar regime.  In Appendix \ref{NumericalProcedure} we give the details
of our numerical procedure. In Appendix \ref{WavesKinetics} we review the
derivation of the kinetic equation for a system of weakly interacting
classical waves.


\section{The Simplest Model of reheating. Numerical Results.}
\label{NumSim:One-Field}

We start with a presentation of our numerical results for the inflaton decay
and the subsequent equilibration of the decay products in a simple $\lambda
\Phi^4$ model. The results were already briefly reported in
Ref. \cite{Micha:2002ey}. The numerical procedure itself is described in
Appendix \ref{NumericalProcedure}. At the end of the Section we will discuss
some expected differences with more complicated models. This order of
presentation allows us to introduce the typical behavior in the systems under
consideration and to formulate concepts and problems.  This will be useful in
the discussion of the general theory of turbulent thermalization, which we
carry out in the following Section.  Further numerical results, obtained for
the simplest $\lambda \Phi^4$ model, and numerical results obtained for more
complicated multi-field systems, will be presented in Sections~\ref{TwoFields}
and~\ref{CompLatt2Kin}.

\subsection{Results for the $\Phi^{4} $-Model}

In this simple model, the field $\Phi$ is the only dynamical variable.  Its
initial homogeneous mode drives inflation, while development and growth of
fluctuations on sub-horizon scales at the end of inflation can be viewed as a
simple model of reheating.  Inflation ends when the motion of the homogeneous
mode of the field changes from the regime of ``slow-roll'' to the regime of
oscillations.  It is convenient to work in conformal coordinates where the
metric takes the form
\begin{equation}\label{metric}
ds^2 = a(\eta)^2\, (d\eta^2 - d\bm{x}^2).  
\end{equation}
We choose the case of a massless field where the equation of motion 
for the rescaled field $ \varphi \equiv \Phi a$ after
inflation is the same as in flat space-time
\begin{equation}\label{BEq0}
\Box \varphi + \lambda \varphi^3 = 0  \,  .
\end{equation}
Therefore, all results obtained in this model are equally applicable to the
reheating of the Universe after inflation and to modeling of other processes
of thermalization in relativistic systems, say, after heavy ion collisions.

The homogeneous component of the field, which corresponds to the zero momentum
in the Fourier decomposition, $\varphi_0 (\eta) \equiv \langle \varphi
\rangle$, is usually referred to as the ``zero-mode.''  It is convenient to
make a rescaling of the field, $\phi \equiv \varphi/\varphi_0(\eta_0)$, and of
the space-time coordinates, $x^\mu \rightarrow \sqrt{\lambda}
\varphi_0(\eta_0) x^\mu$.  Here, $\eta_0$ corresponds to the initial moment of
time (end of inflation). In what follows dimensionless time is still
denoted as $\eta$. With this rescaling, the initial condition for the
zero-mode oscillations is $\phi_0(\eta_0 ) = 1$, and the equation of motion
takes the simple parameter free form
\begin{equation}
\Box \phi + \phi^3 = 0  \,  .
\label{BEq}
\end{equation}
All model dependence on the
coupling constant $\lambda$ and on the initial amplitude of the field
oscillations is encoded now  in the initial conditions for the small
(vacuum) fluctuations of the field with non-zero momenta 
(see \cite{Khlebnikov:1996mc} and appendix \ref{NumericalProcedure}).
The physical normalization of the inflationary model corresponds to
a dimensionful initial amplitude of $\varphi_0(\eta_0) \approx 0.3 M_{\rm
  Pl}$ and a coupling constant $\lambda \sim 10^{-13}$ \cite{lindebook}.
The re-parameterization property of the system allows to choose a larger
value of $\lambda$ for numerical simulations. We have used
$\lambda=10^{-8}$.

Various quantities can be measured in a lattice calculation and monitored as
functions of time.  Here we will discuss the zero mode, $\phi_0
\equiv \langle\phi\rangle$, the variance, $var(\phi) \equiv
\langle\phi^2\rangle -\phi_0^2$ and "particle occupation" numbers.  For
definitions see Appendix~\ref{NumericalProcedure}.

\begin{figure}[t]
  \includegraphics[width=3.8in]{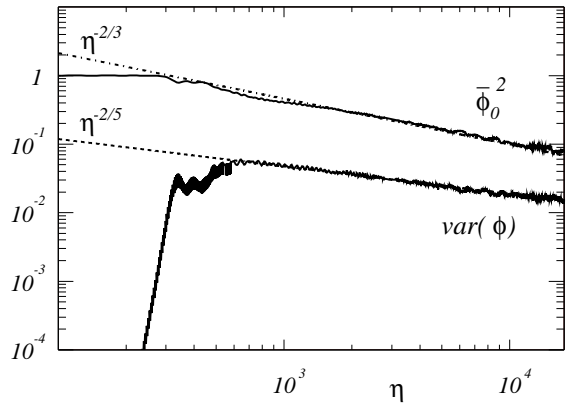}
  \caption{
    Squared amplitude of the zero-mode oscillations, 
    $\overline\phi_0^2$, and variance of the field fluctuations
    as functions of time $\eta$.
    }
  \label{spat_av}
\end{figure}

We begin the discussion of our numerical results with the evolution of the
zero-mode and the variance of the field, which are shown in
Fig.~\ref{spat_av}. The zero mode is a rapidly oscillating function on the time
scale of our lattice calculation.  In Fig.~ \ref{spat_av} we show the amplitude
of oscillations, $\overline\phi_0$, as a function of time.

The initial fast transfer of the zero-mode energy into fluctuations during
preheating (up to $\eta\sim 300$) is followed by a long and slow relaxation
phase. In this late time regime the amplitude of the zero mode oscillations
decreases according to $\sim \eta^{-z}$ with $z \approx 1/3$, the variance of
the field (averaged over high-frequency oscillations) drops according to $\sim
\eta^{-v}$ with $v \approx 2/5$.  In addition, we find that in this regime the
zero-mode is in a non-trivial dynamical equilibrium with the bath of highly
occupied modes: when the zero-mode is artificially removed, it is recreated on
a short time-scale (Bose condensation).

\begin{figure}[t]
\includegraphics*[width=3.9in]{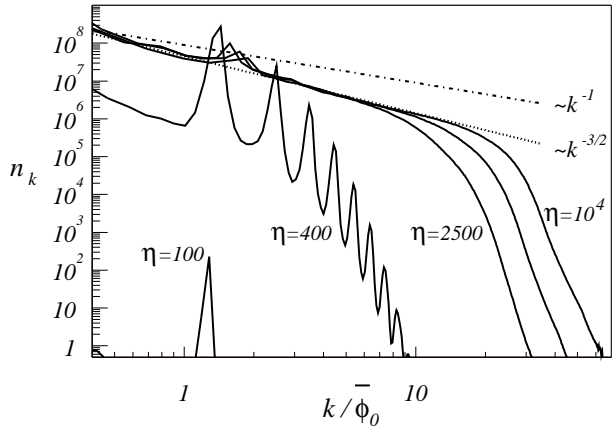}
\caption{Occupation numbers as function of $k/\overline\phi_0$ at\\
$\eta = 100, 400, 2500, 5000, 10000$.}
\label{spectra}
\end{figure}

A detailed analytical discussion of the initial linear stage of the parametric
resonance in this model can be found e.g. in Refs.
\cite{Kaiser:1997mp,Greene:1997fu,Kaiser:1998hg}. During this stage the
occupation numbers grow exponentially with time in a narrow band of resonance
momenta.  Figure~\ref{spectra} shows the occupation numbers at different
moments of time. The displayed spectrum at time $\eta =100$ corresponds to the
stage of parametric resonance. The resonance peak is located at the
theoretically predicted value of $k_{\mathrm{res.}} \sim 1.27$
\cite{Greene:1997fu}. At later time, the growth of the resonance peak is
stopped by re-scattering of particles out of the resonance band, which leads
to a broadening of the occupied region and to the appearance of multiple peaks
\cite{Khlebnikov:1996mc} of comparable width, see spectrum at $\eta
=400$. This structure fits estimates for the development of turbulence in the
presence of a narrow width source located at a finite $k$, see
Ref. \cite{ZakBook}. At even later times the spectra have become smooth
because of re-scattering, and only the first peak is still visible as a small
bump. With time, its position moves towards smaller momenta, reflecting the
change in the effective frequency of inflaton oscillations.  However, if the
particle momenta are rescaled by the current amplitude $\bar\phi_{0} $ of the
zero mode oscillations, as in Fig.~\ref{spectra}, the position of the
resonance is approximately unchanged.  Particles with small momenta are
distributed according to a power law, which at larger momenta is bounded by
a cut-off.  The position of this cut-off moves with time to the
"ultra-violet". This reflects a general tendency of the system to thermal
equilibrium. Indeed, in a state of thermal equilibrium the energy of the
system should be concentrated at much higher wave-numbers compared to the
resonance momenta. On the other hand, energy is inputted into the system of
particles in the region of $k$ near the resonance peak. Therefore, we have a
continuous flux of energy across momentum space, from low to high momenta.

\begin{figure}[t]
\includegraphics*[width=3.9in]{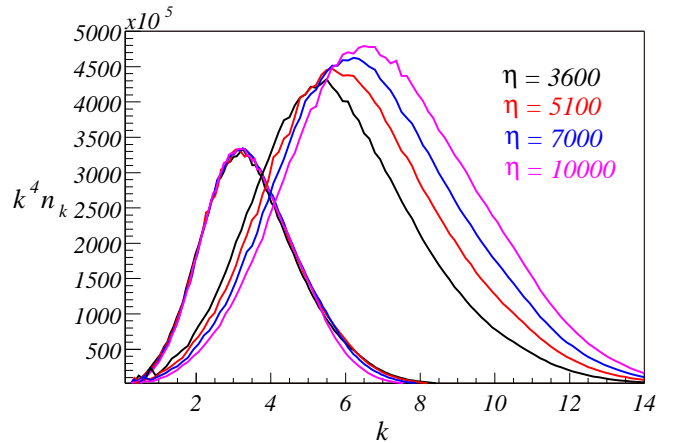}
\caption{On the right hand side we plot
    the wave energy per decade found in lattice integration
    at  $\eta = 3600, 5100, 7000, 10000$. On the
    left hand side are the same graphs transformed according to
    the relation inverse to Eq.~(\ref{SelfS}).
\label{self_sim}}
\end{figure}

This stage of evolution ($\eta>1500$) has the following 
characteristic features :

\begin{enumerate}
\item The system overall is statistically close to a Gaussian distribution of
  field amplitudes and conjugated momenta
  \cite{Felder:2000hr,Micha:2002ey}.
\item The spectra in the dynamically important region can be described by a
power law, $k^{-s}$ with $s \approx 3/2 $.  We see that the system is not in a
thermal equilibrium which would correspond to $s = 1 $. Rather, the exponent
of particle distributions in the power law region corresponds to 
Kolmogorov turbulence \cite{Micha:2002ey}.
\item The power law is followed by a cut-off at higher $k$.  Energy
accumulated in particles is concentrated in the region were the cut-off
starts. Its position is monotonously growing toward the ultra-violet,
reflecting the evolution towards thermal equilibrium.
\item This motion can be described as a self-similar evolution
\cite{Micha:2002ey}
\begin{equation}
n(k,\tau) = \tau^{-q} n_0 (k \tau^{-p}) \, ,
\label{SelfS}
\end{equation}
where $\tau \equiv \eta/\eta_c$ and $\eta_c$ is some (arbitrary) late-time
moment. The best numerical fit corresponds to $q \approx 3.5p$ and $p
\approx1/5$, and is presented in Fig. \ref{self_sim}. The value of the exponent
$p$ is of prime interest since it determines the rate with which system
approaches equilibrium.
\end{enumerate} 
The first and the second point in this list facilitate the use of wave
kinetic theory, see e.g. \cite{zak:85,ZakBook}.  
However,  a straightforward application is difficult and may be even
inappropriate, at least at the early re-scattering stages, because
of the following:
\begin{enumerate}
\item The zero mode does not decay completely. It may induce ``anomalous''  
terms in the collision integral, which are absent in the usual kinetic
description. 
\item The occupation numbers are large initially, of order of the inverse
coupling constant, $n_k \sim 1/\lambda$, see Fig.~\ref{spectra}.  Therefore,
in addition to lowest order collisions (e.g. scattering of two particles into
two particles with different momenta), multi-particle collisions may be
dynamically important as well.
\end{enumerate} 
Therefore, precise lattice calculations are needed. On the other hand, they
have a limited dynamical range in momenta and in time, and one has to switch to
a kinetic approach at some later stage. To determine when (and if) this is
possible, the results obtained with the use of a simple kinetic approach should
be confronted with the lattice results.

In the following Sections we will develop and apply the theory of weak wave
turbulence to the models of the type integrated on the lattice. In
particular, we will calculate all universal scaling exponents and show that
they are in  agreement with lattice results. At ``early'' times the
dynamics of the model described above is driven by $m$-particle interactions
with $m=3$. Wave turbulence theory gives for scaling exponents in $d=3$
spatial dimensions:
\begin{eqnarray}
&& p = 1/(2m-1)\; , \nonumber \\
&& s = d - m/(m-1)\; , \nonumber \\
&& v = 2/(2m-1)\; , \nonumber \\
&& z = 1/(d(m-1)-m)\; . \nonumber
\end{eqnarray}
 
\subsection{Expected differences in more complicated models}
\label{Preview:other models}

The flux of energy over momentum space, which is necessarily present in
problems like reheating and thermalization after inflation, signifies that we
should observe a turbulent state during the thermalization stage and that
the theory of turbulence applies.  In a simple $\lambda \phi^4$ model the
stage of preheating (i.e. parametric resonance) ends when roughly half of the
inflaton energy is transferred to particles. Indeed, Fig.~\ref{spat_av} shows
that the amplitude of the zero mode, which is a source of energy for the
turbulence problem, starts to decrease already at the end of the parametric
resonance stage. In such system we expect the free turbulence regime to
follow the preheating stage. 

In more complicated systems, which involve other fields coupled to the
inflaton, say, some field $\chi$, parametric resonance may end when the
fraction of energy transferred to the $\chi$ excitations is still negligible
compared to the energy stored in the inflaton zero mode. Indeed, parametric
resonance ends when the rate of re-scattering of particles out of the
resonance band became comparable to the resonant production rate and the
maximal value of the variance of $\chi$ excitations achieved at the end of the
resonance stage is $\sim 1/g^2$, where $g^2$ is either the coupling of $\chi$
to the inflaton, or self-coupling of $\chi$ (viz., the largest of these
two). We expect that in this case turbulent transport will develop when the
amplitude of the inflaton zero mode is still unchanging.  This means, that the
transfer of zero mode energy into $\chi$-filed should occur in the regime of
stationary turbulence.  Only when the amount of energy in the zero-mode
becomes subdominant we should expect a transition to the regime of free
turbulence.  This is an important difference to the simple $\phi^{4} $-model.
In particular, the distribution functions move much faster into the ultra-violet in
this regime, $p = (m-1)/(2m-1)$.  We will see that the regime of stationary
turbulence is indeed present in two field models, see Section~\ref{TwoFields}.

\section{Thermalization in the Wave Kinetic Regime. General Theory.}
\label{Turbulentreheating}

\subsection{Turbulent reheating: a motivation}

Kolmogorov's turbulence is characterized by a stationary transport of some
conserved quantity between different scales in momentum (Fourier) space
\cite{K41:1,K41:2}.  In the following, we will restrict ourselves to systems
with spatially isotropic and homogeneous correlation functions, which applies
to the cosmological conditions after inflation.  Turbulence usually appears
when a source of energy or particles is present and is localized in some
momentum region $k_{\mathrm{in}}$. In addition to the source exists a "sink"
localized at $k_{\mathrm{out}}$. When both, source and sink are stationary, it
is natural to expect the eventual development of a stationary state with scale
independent transport of the conserved quantity through momentum
space. Indeed, energy or particle number cannot accumulate between
$k_{\mathrm{in}}$ and $k_{\mathrm{out}}$ and should flow from one scale to the
other.

This is a system-independent formulation of Kolmogorov's concept of
turbulence, which he formulated in the context of hydrodynamical systems
\cite{K41:1,K41:2}.  Zakharov applied it to systems of coupled waves
\cite{Z67} in the regime of kinetic wave interactions.  His approach is based
on his derivation of the wave kinetic equations (see
e.g. \cite{Z67,zak:85,ZakBook}) and is well suited to studies of turbulence in
classical field theories. We will adopt it here.

The physical scenario of reheating after inflation shares basic ingredients
with that of turbulence: there exists a localized source of energy- the
coherently oscillating inflaton zero-mode - pumping energy into the system of
particles at Fourier wave-numbers $k_{\mathrm{in}} \sim k_{\mathrm{res}}$.
The mechanism behind this pumping can be parametric resonance, tachyonic
amplification, etc.  Like in the turbulent scenario there do not exist other
intermediate scales (wave-numbers), where energy is infused, accumulated, or
dissipated.  Thus, it seems likely that the eventual dynamics of reheating -
after the explosive regime of preheating has ended - is close to that of
Kolmogorov's turbulence.

However, in the description of reheating appear some differences to
stationary turbulence, since:
\begin{enumerate}
\item A sink does not exists.
\item The source (i.e. the amplitude of inflaton zero-mode oscillations and
  therefore interaction rates) can be essentially time-dependent on  relevant
  time scales.
\item Neither source nor sink exist when the inflaton has completely decayed.
\end{enumerate}
 In the first case, we expect that the stationary turbulent flux of energy
still will be established in some ``inertial'' range $k_{\mathrm{in}} < k <
k_{\mathrm{out}}$. Particle distributions in this range of momenta should not
be significantly different compared to the case with a stationary
sink. Indeed, in the typical turbulent problem the energy dissipates
(e.g. into heat) after entering the region $k \gg k_{\mathrm{out}}$.  For
problems relevant to thermalization after inflation, instead of dissipation
the transported energy is used to populate high momentum modes at $k \gg
k_{\mathrm{out}}$.  If the transport is reasonably ``local'' in momentum
space, the flux of energy through the inertial range should not be influenced
much by processes which involve $k > k_{\mathrm{out}}$. Energy may dissipate
at $k_{\mathrm{out}}$ or continue the flow to even higher momenta, but
regardless of this, we should expect the same distribution of particles in the
inertial range. However, in the latter case we can expect that the value of
$k_{\mathrm{out}}$ increases, and since the flux of energy is constant
throughout inertial range, the total energy of a system without a sink has to
grow linearly with time,
\begin{equation}\label{Epropt}
E(t) \propto t \; .
\end{equation}
This is a simple consequence of the stationarity of turbulence in the
inertial range, and can be used as its signature.

A time dependent source (second point above) changes the picture somewhat,
since stationary states are not likely to develop even in a finite range of
$k$.  However, a weak time-dependence should still allow for a
close-to-stationary and close-to-turbulent evolution.  Moreover, even if the
source eventually does not exists, particle distributions in the inertial
range as functions of momenta can still be close to turbulent power
laws. Indeed, stationary turbulent distributions can be found as zeros of the
collision integral \cite{ZakBook}. In the non-stationary case the collision
integral is non-zero, but should approach a minimal value in the
inertial range which may result in the same shape of particle distributions
there.  

\subsection{Wave turbulence by scaling analysis}

The dynamics of coupled waves close to a stationary state can be described by
a wave kinetic equation (see e.g. refs.~\cite{ZakBook,Z67,Newell:2001}):
\begin{equation}\label{genkineq}
\dot n_{k} = I_{k}[n]\;.
\end{equation} 
Here the function $n_{k}$, usually called \textit{occupation number} or
\textit{wave action}, describes the average volume of phase space occupied by
the oscillations of a single mode with a wave-number $k$.  Its evolution is a
result of resonant wave interactions, the effect of which is described by the
collision integral $I_{k}[n]$.  The collision integral is a function of the
``external'' momentum $k$ and a functional of the distribution function $n$,
which is reflected in the notations we use. When we do not need to stress the
functional dependence, we will also write $I_k$ as $I(k)$. The collision
integral for the case of interest, Eq. (\ref{BEq0}), is explicitly derived in
Appendix \ref{WavesKinetics}.

Before we proceed, let us remind the general structure of the collision
integrals using as illustration the scattering of two particles into two
particles, which will be referred to as 4-particle process. This will also
allow us to introduce the necessary notations.  In all cases we will write the
collision integral as
\begin{equation}\label{coll-int:def1}
I_k[n] = \int d\Omega (k,q_i)\, F(k,q_i) \, .
\end{equation}  
This form separates the contributions which are due to the (fixed) particle
model, $d\Omega(k,q_i)$, from those which are due to the (evolving) particle
distribution functions, $F(k,q_i)$. Here $k$ is the external momentum and
$q_i$ refer to momenta over which the integration is carried out.  If $m$
particles participate in the collision, $i$ takes values from 1 to
$m-1$. E.g. when 2 particles scatter into 2 particles, $m=4$ and there are 3
internal momenta over which we integrate, $q_1$, $q_2$ and $q_3$.  Namely
\begin{equation}\label{Omega-def}
d\Omega (k,q_i) = 
\frac{(2\pi)^4 |M|^2}{2 \omega_k}
\delta^4(k_\mu,q_{i\mu}) 
\prod_{i=1}^3
\frac{d^3{q}_i}{2 \omega_i (2\pi)^3 } .
\end{equation}
$d\Omega$ contains the usual energy-momentum conservation $\delta$-functions,
which we have denoted as $\delta^4(k_\mu,q_{i\mu})$, the ``matrix element''
squared, $|M|^2$, of the corresponding process (which is a function of $k$ and
$q_i$) and the integration measure over momentum space. Here, $k_0 = \omega_k
= \omega(k)$ and $\omega_i = \omega(q_i)$ refers to the particle energy. 

When quantum effects are accounted for, the function $F$ in our example is
given by
\begin{eqnarray}
F(k,q_i) &=& (1+n_k)\,(1+n_{q_1}) n_{q_2} n_{q_3} \nonumber \\ 
&-& n_k n_{q_1} (1+n_{q_2})\,(1+n_{q_3}) \; .
\label{FDef}
\end{eqnarray}
In the limit $n \gg 1$ terms $O(n^2)$ can be neglected and
$F$ is a sum of terms $O(n^3)$
\begin{equation}
F(k,q_i) = 
(n_k + n_{q_1}) n_{q_2} n_{q_3}
- n_k n_{q_1} (n_{q_2} + n_{q_3})\; .
\label{FLargeN}
\end{equation}
The limit $n \gg 1$ corresponds to interaction of classical waves and
expression Eq. (\ref{FLargeN}) is also explicitly derived in Appendix
\ref{WavesKinetics}.  This illustrates a general rule: in the classical limit
and for interaction of m waves the function $F$ is a sum of terms $O(n^{m-1})$
with appropriate permutations of signs and indices.  In other words, in this
limit $F$ is a homogeneous function with respect to multiplication of each
occupation number by $\zeta$
\begin{equation}\label{F-scaling}
F(\zeta n) = \zeta^{m-1} F(\zeta n) \; .
\end{equation}
This property is extremely important in our subsequent analysis.  When quantum
effects become important (i.e. when one should properly write $[1+n]$ in F),
the classical turbulence and/or self-similar evolution stops. At that moment
particle distributions relax to usual Bose-Einstein functions. We will not be
concerned here with the (presumably relatively short) relaxation period from
the classical to the quantum regime, but will study in detail the turbulent
evolution in the regime of classical waves.

This gives us sufficient notational details to proceed with the discussion of
turbulence.  We restrict it to systems which are isotropic and homogeneous in
configuration space, when occupation numbers (as well as all other parameters
which enter the collision integral) depend on the modulus of momenta only.  We
consider the classical limit in the function $F$ with general m-particle
interaction, in case of which Eq.~(\ref{F-scaling}) holds.  To keep the
discussion general, in the rest of this section we will consider the case of
(d+1) dimensional space time.

Often a collision integral conserves one or several quantities. We restrict
ourselves to energy density
\begin{equation}\label{rho:def}
\mathcal{\rho}=\int \frac{d^dk}{(2\pi)^d}\,
\omega_{k}\,n_{k}  \;, 
\end{equation}
which is conserved when the expansion of the Universe can be neglected or
``rotated'' away, and particle density
\begin{equation}\label{n:def}
{n}=\int \frac{d^dk}{(2\pi)^d}\,n_{k}
\;,
\end{equation} 
which corresponds to conserved charges, e.g. baryon number.

Conservation of $n$ or $\rho$ can be expressed as a continuity equation in 
Fourier space, e.g.
\begin{equation}\label{continuity}
\partial_t (\omega_{k}\,n_{k}) + \nabla_k \cdot j_k =0.  
\end{equation}
Here and in what follows we will write the explicit relation for energy
conservation, the case of conserved charges can be easily obtained by a formal
substitution $\omega_{k} = 1$.  In the isotropic case only the radial
component of the flux density, $j_k$, is non-vanishing and we get for the
energy flux, ${S}^{\rho}(p)$, trough the sphere of radius $p$
\begin{equation}
\begin{split}
{(2\pi)^d}\cdot {S}^{{\rho}}(p) =& 
- \int^p {d^dk} \omega_k\, \dot n_{k}\\ =& 
- \frac{\pi^{d/2}}{\Gamma(1+\frac{d}{2})}\int^p dk k^{d-1} \omega_k\, I_{k}[n]
\; \, ,
\end{split}
\label{Frho_def}
\end{equation} 
In \eqref{Frho_def} the factor in front of the integral is the area of the
d-dimensional unit sphere.  In case of stationary turbulence this flux should
be scale-independent, i.e. integral Eq. \eqref{Frho_def} should not depend
upon its integration limit $p$. This is possible if the collision integral
equals zero. One can explicitly look for solutions $ I_{k}[n] = 0$, see
e.g. \cite{ZakBook}. Such solutions correspond to stationary turbulence and
exist with non-trivial boundary conditions (source and sink), in addition to
the Rayleigh-Jeans-law of classical equilibrium. Here we adopt an alternative
and somewhat simpler approach of Ref.~\cite{kats:76} to determine the
turbulent solutions.

Following \cite{kats:76} we consider states for which the collision integral
has certain scaling properties under $\xi$-rescaling of the external momentum
$k$
\begin{equation}\label{nu-scal}
I_{\xi k}[n] = \xi^{-\nu} I_{k}[n] \;.  
\end{equation}
To simplify notations we assume that all momenta were made dimensionless by
rescaling with some typical momentum scale, without explicitly writing this.
The special choice $\xi=k^{-1}$ allows us to find the $k$-dependence of the
collision integral, $I_{k}[n] = k^{-\nu} I_{1}[n]$.  Let us additionally
assume that the dispersion law is a homogeneous function as well,
\begin{equation}\label{omega-scal}
\omega (\xi k) = \xi^{\alpha}  \omega (k)\;.  
\end{equation}
Relations \eqref{nu-scal}and \eqref{omega-scal} should hold in some region of
momenta where we expect turbulent behavior. Integrating Eq. (\ref{Frho_def})
we find
\begin{equation}\label{F-scal-k}
{S}(p)\;\propto\;-\;p^{d+\alpha-\nu} 
\frac{I_{1}(\nu )}{d+\alpha-\nu} 
\end{equation} 
Here we indicated explicitly that the collision integral in the turbulent
state with scaling behavior Eq. (\ref{nu-scal}) depends on the exponent $\nu$.
We find that the flux is scale invariant, if
\begin{equation}\label{t-ndex1}
\nu = d+\alpha \; .
\end{equation} 
This condition defines the turbulent exponents which we will specify in detail
below. Note that this implies the existence of the limit
\begin{equation}\label{turbcond:limit}
\lim_{\nu \rightarrow d+\alpha} \;
\frac{I_{1}(\nu)}{d+\alpha-\nu}\;=\;\mathrm{const}\neq 0\;,
\end{equation}  
as a sufficient condition for the existence of a stationary turbulent solution:
if the collision integral has a zero of first degree at $\nu = d+\alpha$ , the
turbulent flux is scale-invariant and finite.

In what follows, we consider particle models for which $d\Omega$ is a
homogeneous function of all momenta
\begin{equation}\label{Omega-scal}
d\Omega (\xi k, \xi q_i) = \xi^{\mu}\,  d\Omega (k,q_i)\;.  
\end{equation}
Rescaling of the external momentum $k$ by $\xi$ gives 
\begin{equation}\label{coll-int:rescale}
I_{\xi k} = \xi^{\mu} \int d\Omega (k,q_i)\, F(\xi k,\xi q_i) \, ,
\end{equation}  
since integration over every $q_i$ is from 0 to $\infty$.  We will exploit
this relation in two ways:

\begin{enumerate}
\item Often the evolution of distribution functions involves rescaling
  of their momenta, see Sec.~\ref{sec:self-sim}. If this is the case, the
  collision integral as a function of time can be found with the help of
\begin{equation}\label{coll-int:use1}
\int d\Omega (k,q_i)\, F(\xi k,\xi q_i) = \xi^{-\mu} \, I_{\xi k}\, .
\end{equation}  
\item Let us assume that the particle distribution functions are power laws in
  the momenta, 
\begin{equation}\label{def:s-exponent}
n(q) \propto q^{-s} \; . 
\end{equation}  
This leads to the following scaling of F
\begin{equation}\label{coll-int:use2}
F(\xi k,\xi q_i) = \xi^{\,- s\, (m-1) } \, F(k,q_i)\; , 
\end{equation} 
\end{enumerate} 
Combining this with Eq.~\eqref{coll-int:rescale} 
we find $I_{\xi k} = \xi^{\mu - s\, (m-1)}\, I_k$. A comparison with
Eqs.~(\ref{nu-scal}) and (\ref{t-ndex1}) leads us to the exponent $s$ which
defines the scaling of particle distribution functions in a turbulent
state with constant energy transport (we will call this {\it energy
  cascade} for brevity)
\begin{equation}\label{t-constE}
s = \frac{d + \alpha + \mu}{m-1} \; .
\end{equation}  
Turbulence with constant transport of particle number (similarly, we will
call this state {\it particle cascade}) can be found at this point by the
formal substitution $\omega = 1$, i.e. $\alpha = 0$ and
\begin{equation}\label{t-constN}
s = \frac{d + \mu}{m-1} \; .
\end{equation}  
Note that doing this substitution at later stages would be confusing since the
explicit expression for $\mu$ also contains $\alpha$.  Note also that on
turbulent states $I[n] = 0$, therefore, transport of all quantities except
energy is zero for energy cascade. For particle cascade, which describes
Bose-condensation \cite{Semikoz:1995zp,Semikoz:1997rd}, the transport of
energy is zero.

The reader should bear in mind that only those solutions that describe the
transport of energy towards the ultra-violet, $ {S}^{{\rho}} > 0$, are
relevant for the problem of thermalization after inflation. The sign of fluxes
for stationary turbulence of three- and four-wave collision integrals was
found in Ref.~\cite{kats:76},
\begin{equation}\label{sign-of-S}
{\rm sign}\, {S}^{{\rho}} = {\rm sign}\,[ \alpha s (s-\alpha)] \; .
\end{equation}  
In thermal equilibrium $n \propto \omega^{-1}$, i.e. $s = \alpha$. Therefore,
energy turbulence is directed towards the ultraviolet if the distribution
function with increasing momenta falls off faster than in equilibrium, $s >
\alpha$. As we will see, in the $\lambda \phi^4$ model this condition holds in
$d=3$, but is violated at $d \leq 2$. Therefore, we believe that simulations
of the thermalization in this model at $d < 3$, see
e.g. Refs. \cite{Aarts:2000mg,Borsanyi:2000ua, Borsanyi:2000pm,
Boyanovsky:2003tc}, may not reflect all aspects of the physical problem of
reheating after inflation correctly.

\subsection{Self-similar evolution}
\label{sec:self-sim}
In an analytical approach to non-stationary situations (e.g. when describing
free turbulence) it is usually assumed that the evolution is self-similar
\cite{Falk:1991,zak}.  As we have shown, the evolution is self-similar,
indeed, at late times in our numerical integration of the $\phi^{4} $-model,
see Section~\ref{NumSim:One-Field}.  Below we consider self-similar
substitutions in anticipation that they provide a valid leading description of
thermalization in the class of models we consider.
 
Let $n_0(k)$ be a distribution function at some late moment of time $t_0$,
when the regime of self-similarity has been already established.  The
subsequent evolution can be described as rescaling of momenta accompanied by a
suitable change of the overall normalization
\begin{equation}\label{ss1} 
n(k,\tau) = A^{\gamma}\,n_0(kA)\;,
\end{equation}
where we have defined $\tau \equiv t/t_0$, $\gamma$ is some constant and $A =
A(\tau)$ is some time dependent function satisfying $A(1) = 1$. Both,
$A(\tau)$ and $\gamma$, are determined by the solution of the kinetic
equation (\ref{genkineq}).

In some cases the collision integral may contain an additional explicit time
dependence which can be isolated as an overall factor $B(\tau)$. This factor
may be induced by time-dependent classical backgrounds like the scale factor
of the expanding universe or the zero-mode of the inflaton field.  It is
convenient to rescale the collision integral by some typical rate $\Gamma$, $I
\equiv B \Gamma \tilde{I}$, such that $B$ and $\tilde{I}$ are
dimensionless. We use $B(1) = 1$ as normalization.

When Eq. (\ref{F-scaling}) holds, the factor $A^{\gamma}$ of each distribution
function, Eq. (\ref{ss1}), can simply be taken out of $F$ and out of the
collision integral, which becomes a functional of $n_0$. After that we can use
Eq.  (\ref{coll-int:use1}) with $\xi = A$ which gives
\begin{equation}\label{ci-ss} 
I(k,\tau) = A^{\gamma(m-1)-\mu} B \Gamma \tilde{I}_{kA}[n_0]\;,
\end{equation} 
On the other hand, the l.h.s. of the kinetic equation (\ref{genkineq}) can be
written as
\begin{equation}\label{lhs-ss} 
\dot{n}(k,\tau) = A^{\gamma-1} \dot{A} \left( \gamma n_0 + \zeta 
\frac{d n_0}{d \zeta} \right) \; , 
\end{equation} 
where we have defined $\zeta \equiv k A$. Using $\Gamma$ as
a separation constant, the kinetic equation can be split into two:
one for the shape of the distribution function,
\begin{equation}\label{eq4n} 
\gamma n_0 + \zeta \frac{d n_0}{d \zeta} = - \tilde{I}(\zeta ) \; ,
\end{equation} 
and one for the dynamical evolution
\begin{equation}\label{eq4A}
A^{\mu - \gamma (m-2) -1}\frac{dA}{d\tau} =  - \Gamma t_0 B\; .
\end{equation} 
We will not be concerned with \eqref{eq4n}   here and simply assume that it has
some non-trivial solution. 
The general solution of \eqref{eq4A} is of the  form  
\begin{equation}\label{A-gSol}
A = \Theta^{-p} \; ,
\end{equation} 
where
\begin{equation}\label{tautilde} 
 \Theta \equiv \frac{\Gamma t_0}{p}\int_1^\tau B(\tau')d\tau'  +1 
\end{equation} 
and
\begin{equation}\label{p-Defs}
p \equiv \frac{1}{\gamma (m-2) - \mu} \; .
\end{equation} 
We fix scales using the condition $\Gamma t_{0} =p $.  For a time-independent
background $ B $, i.e. $B\equiv 1$, it than follows, that $\Theta =\tau $ and
eq. \eqref{A-gSol} simplifies to
\begin{equation} \label{A-gSol-simple}
A=\tau^{-p}\;.
\end{equation} 
We will discuss this case first. 

\subsubsection{Self-similar evolution in time-independent background}

Substituting \eqref{A-gSol-simple} in \eqref{ss1} we obtain
\begin{equation}\label{ss2} 
n(k,\tau) = \tau^{-\gamma p}\,n_0(k \tau^{-p})\;,
\end{equation}
In applications of turbulence theory to thermalization, this solution is most
important.  Let $k_c$ be the initial value of some characteristic momentum
scale, e.g. the scale where most of the energy carried out by a self-similar
distribution is concentrated. According to Eq.~(\ref{ss2}), with time this
scale evolves as
\begin{equation}\label{k_c-motion}
k_c(\tau ) = k_c(1)\; \tau^p  \, .
\end{equation}
The exponent $p$ determines the speed with which the distribution function
moves over momentum space and therefore defines e.g. the time scale of
thermalization.  This is a reason why we will be interested mainly in the
value of the exponent $p$, ~Eq.~(\ref{p-Defs}).  In applications to
thermalization after preheating the energy is concentrated at low momenta
initially and should propagate to high momenta. This means that solution
Eq.~(\ref{ss2}) is physically relevant for $p > 0$.

The exponent $\gamma$, which enters Eq.~(\ref{p-Defs}) can be fixed by
specifying appropriate boundary conditions, which are specified below.

\paragraph{Isolated systems}
\label{isolated} 
If the wave energy is strictly conserved it follows that
\begin{equation}\label{EnCons}
\begin{split}
{\rm const } &= \int d^d k\, \omega_{k}\, n(k,\tau)\\ &= 
A^{\gamma - (d+\alpha )} \int d^d \zeta\, \omega_{\zeta}\, n_0(\zeta ) \; .
\end{split}
\end{equation} 
This gives
\begin{equation}\label{gamma-EnCons}
\gamma = d+\alpha \; .
\end{equation} 
Similarly, for the evolution with particle number
conservation one obtains $\gamma = d$. Here we would like to stress the
following subtlety. Clearly, a simple self-similar substitution
Eq.~\eqref{ss1} cannot account for energy and particle number conservation
simultaneously, while both quantities are conserved in a number of systems. If
this is the case, one should choose the integral which gives dominant
restriction of $n_k$, i.e. the energy for energy cascade (thermalization) and
particle number for the inverse cascade (Bose-condensation). For the problem of
thermalization of ultra-relativistic particles this gives
\begin{equation}\label{p-EnCons-rel}
p_{i} = \frac{1}{(d+1)(m-2) - \mu}~~~~~~~~~
\begin{aligned}  
&{\rm relativistic}\\
&{\rm energy~ cascade} 
\end{aligned}
\end{equation} 
However, describing thermalization in the non-relativistic limit, $\omega_k =
m + k^2/2m$ with $k^2/m^2 \ll 1$, we can neglect the kinetic energy with
respect to the rest mass in the normalization condition \eqref{EnCons},
i.e. we should use $\gamma = d$, as in the case of particle conservation
\begin{equation}\label{p-EnCons-non-rel}
p_{i} = \frac{1}{d (m-2) - \mu}~~~~~~~~~
\begin{aligned}  
&{\rm non-relativistic}\\
&{\rm energy~~ cascade} 
\end{aligned}
\end{equation} 

\paragraph{Driven turbulence.}
\label{driven}
In our lattice integrations we have found that particle distributions as
functions of $k$ follow a power-law in the wake of a propagating energy
 front, $n_{k}(\tau) =\left(b(\tau)/k\right)^{s}$, with exponent $s$
being in agreement with the theoretical predictions for stationary
turbulence. Such behavior is expected \cite{Falk:1991} for the regime of
driven turbulence in the presence of a stationary source (and then $b(\tau) =
{\rm const}$). However, for the case of free turbulence we are not aware of
any predictions. Here we consider consequences of such a behavior assuming
general $b(\tau)$ (the case of constant $b$ being a particular case).

Considering distribution the function in the region of low momenta,
$n_{k}(\tau) =(b/k)^{s} = A^{\gamma}\,n_0(kA)$ we find
\begin{equation}\label{renorm-spectra}
b \propto A^{{\gamma}/{s}-1}=\tau^{(1-{\gamma}/{s})p}   \;,
\end{equation} 
i.e. the transport of energy through the inertial range is stationary if
\begin{equation}\label{gamma-stationary}
\gamma = s \; .
\end{equation} 
This generalizes the concept of stationary turbulence to a
system without sink.  (Notice, that this requires a stationary source in the
infra-red.) In this regime the total energy in particles has to grow linearly
with time.  Considering the r.h.s. of relation \eqref{EnCons} with $\gamma =
s$ we find $\tau = A^{\gamma - (d+\alpha)} = \tau^{p(d+\alpha -s)}$, or
\begin{eqnarray}\label{p-turb3}
p_t = 1/(d +\alpha -s) \; ,
\end{eqnarray}
where we denote the exponent $p$ for the case of a stationary transport as
$p_t$ to distinguish it from the exponent which corresponds to an isolated
system, $p_i$.  Substituting explicitly the exponent $s$ of the spectra of
stationary turbulence, Eq.~(\ref{t-constE}), we find
\begin{eqnarray}\label{p-turb2}
p_{t} = \frac{(m-1)}{(d+\alpha)(m-2) - \mu} 
= (m -1)\, p_{i} \; .
\end{eqnarray}
The latter relation could have been also found using Eq.~(\ref{t-constE}) and
Eqs.~\eqref{p-Defs} with $\gamma=s$.

\paragraph{Non-stationary source.}
\label{non-stationary}
Let us consider the somewhat more general situation and assume that the energy
inputted into (or taken out from) the system of particles changes with time as
$E(\tau) = E_0 \tau^r$. Clearly, the isolated system corresponds to $r=0$,
while a stationary source corresponds to $r=1$.  We will now have $\gamma =
(d+\alpha ) - r/p $ ~and
\begin{equation}\label{p-nonstat}
p = \frac{1 + r(m-2)}{(d+\alpha)(m-2) - \mu} \; .
\end{equation}

\subsubsection{Time-dependend background}
\label{sec:ExplTimeDep}

We now consider a time-dependent $B$ in Eqs.~\eqref{A-gSol}, \eqref{tautilde}.
As an illustration we choose $B(\tau) = \tau^{-\kappa}$, which gives
\begin{equation}\label{A-GenSol}
\Theta = \frac{1}{(1-\kappa)} (\tau^{1-\kappa}-1) + 1 \;.	
\end{equation} 
Note, and this is important for the interpretation of our numerical
results, that the linear approximation for small times, $ \tau\sim 1 $,
gives $\Theta \simeq \tau $, which brings us back to the  situation considered
in the previous subsection.

The late time behavior, $\tau \gg 1$, depends on the sign of
$1-\kappa > 0$. If $1-\kappa > 0$, the distribution propagates to the
ultraviolet without bound, $A(\tau) \propto \tau^{-(1-\kappa)p}$ and $k_c(\tau
) \propto \tau^{(1-\kappa) p}$.  In other words, at late times $ A
\sim\tau^{-\tilde{p}} $ with
\begin{equation}\label{A-GenSol1}
\tilde{p} = (1-\kappa) p
\end{equation} 
for any boundary conditions discussed above in paragraphs\ref{isolated} -
\ref{non-stationary}.

However, $A(\tau )$ approaches a finite limit at $\tau \rightarrow \infty$ if
$1-\kappa < 0$
\begin{equation}\label{A-GenSol2}
 A(\tau = \infty) = \left[ 1 + \frac{1}{\kappa - 1} \right]^{-p}
 \; .
\end{equation} 
The propagation of particle distribution functions towards the
ultraviolet is limited. This has important consequences for the
thermalization of massive particles in the expanding Universe, as we
shall discuss in more detail below.
	
Expressions Eqs.~\eqref{p-EnCons-rel}, \eqref{p-turb2} and \eqref{A-GenSol1}
are the main results of this section. They determine the speed of propagation
of the particle distribution in momentum space for a specific models.

\section{Stationary States and Self-Similar Evolution in Specific Models }
\label{sec4}

Here we apply the general results of the previous section to a number of
particular models of interest.

First of all we have to determine the scaling exponent $\mu$ of $d\Omega$ (see
Eq.~(\ref{Omega-scal})).  The scaling of $\omega$ is different in relativistic
and non-relativistic regimes. This is accounted for differently in the
argument of the energy conservation $\delta$-function (where in the
non-relativistic regime $\omega(k)$ is replaced by $k^2/2m$) and in the
$1/\omega$ factors of relativistic integration measure (where $\omega$ is
replaced by $m$).  To make the discussion of relativistic and non-relativistic
cases uniform, we move $\omega$ out from the relativistic integration measure
and define the function $U(k,q_i)$
\begin{equation}\label{U-def}
U(k,q_i) \equiv \frac{(2\pi)^{d+1} |M_k|^2}{2 \omega_k \prod_{i=1}^{m-1} 2
\omega_{q_i}} \; .
\end{equation}
In what follows we will assume that in a dynamically interesting range of wave
numbers $U$ follows a scaling law
\begin{equation}\label{U-scaling}
U(\xi k, \xi q_i) = \xi^{\beta} U(k,q_i)\;
\end{equation}
With this definition
\begin{equation}\label{Omega-def2}
d\Omega (k,q_i) = 
U(k,q_i) \delta^d(k_\mu,q_{i\mu}) 
\prod_{i=1}^{m-1}
\frac{d^d{q}_i}{(2\pi)^d }\; ,
\end{equation}
and we find
\begin{equation}\label{mu-result}
\mu =d(m-2) - \alpha +   \beta \; .
\end{equation}

We calculate the exponents $\mu$, $s$, and $p$  for two classes of models.
The first one is characterized by $k$-independent matrix elements,
the second one has no dimensionful parameters. 
The scalar field models which we integrated on the lattice belong to the first
class. In the absence of a zero mode in the relativistic limit in $(3+1)$
dimensions they belong to the second class as well.

\subsection{Theory with $k$-independent matrix elements}

For models with k-independent matrix elements the scaling of $U$ is determind
by the $\omega$'s, and we have $\beta = -m$ in the relativistic regime and
$\beta = 0$ in the non-relativistic case. Eq.~\eqref{mu-result} gives
\begin{eqnarray}\label{mu-Mkindep-rel}
&&\mu =  d(m-2) - 1 - m ~~~~~~({\rm relativistic})   \; , \\
&&\mu =  d(m-2) - 2  ~~~~~~~~({\rm non-relativistic})  \; .
\label{mu-Mkindep-nonrel}
\end{eqnarray}
Substituting these expressions into Eqs.~\eqref{p-EnCons-rel},
\eqref{p-EnCons-non-rel} we find that in this class of models the exponents
$p$ do not depend on the number of dimensions. In particular, for the energy
cascade in an isolated system we have
\begin{eqnarray}\label{p-Mkindep-rel}
&&p_i =  {1}/{(2m-1)}~~~~~~~~({\rm relativistic})   \; , \\
&&p_i =  {1}/{2}~~~~~~~~~~~~({\rm non-relativistic})  \; .
\label{p-Mkindep-nonrel}
\end{eqnarray}
For $m=3$ and $m=4$ Eq.~\eqref{p-Mkindep-rel} gives $p=1/5$ and $p=1/7$
respectively.

Substituting  Eqs.~\eqref{mu-Mkindep-rel}, \eqref{mu-Mkindep-nonrel} into
Eq.~\eqref{t-constE} we find the exponent $s$
\begin{eqnarray}\label{s-Mkindep-rel}
&&s =  d - \frac{m}{m-1}~~~~~~~~({\rm relativistic})   \; , \\
&&s =  d~~~~~~~~~~~~~~({\rm non-relativistic})  \; .
\label{s-Mkindep-nonrel}
\end{eqnarray}
In the non-relativistic regime both exponents, $p_i$ and $s$ do not depend
on $m$. 

\subsubsection{Three-particle interactions, relativistic regime}
\label{sec:3-part}

Three-particle processes appear in the $\lambda \phi^4$ model when
interactions with the zero-mode are important, see Appendix~\ref{WavesKinetics}
and Section~\ref{sec:zero-mode}.

According to Eq.~\eqref{p-Mkindep-rel} for $m=3$ the front of the energy
cascade propagates with
\begin{equation}\label{p-EnCons-3part}
p_i = \frac{1}{5} \; ,
\end{equation}
regardless of of the number of spatial dimensions, $d$.  For the exponent $s$
of particle distributions in the inertial range in $d=3$ we find
\begin{equation}\label{s-EnCons-3part}
s =  \frac{3}{2} \; .
\end{equation}
Both exponents coincide with what is observed in our numerical
experiments. Note that the exponent $s$ is expected to appear in the case of
driven turbulence. In the case of free turbulence the wake of the propagating
turbulent front does not even have to be a power law. Nevertheless, we do
observe a power law with the exponent $s=3/2$ to a very good accuracy.  This
might be not a chance coincidence. However, in $d < 3$ the theory predicts $s
< 1$, the spectrum falling-off with $k$ more slowly compared to  thermal
equilibrium, and one can get a different shape of particle distributions in
$d<3$ (but we still expect the exponent $p$ to be given by
Eq.~\eqref{p-Mkindep-rel}).

\subsection{Relativistic theory with dimensionless couplings.}

The $\lambda \phi^4$ model in $d=3$ which we have simulated on the lattice
belongs to the class of models considered in this paragraph. In $d=2$
dimensionless couplings appear in  the $\lambda \phi^6$ model. Dimensionless
couplings are generic and this case is not restricted to  scalar field models,
therefore we consider it separately.

If the collision integral does not contain any dimensionfull parameters, it
has to scale with $\mu = 1$ and we find for the exponent $p_i$ of energy
conserving propagation in an isolated system, Eq.~(\ref{p-EnCons-rel})
\begin{equation}\label{p-EnCons-rth}
p_{i} = \frac{1}{(d+1)(m-2) - 1} \; .
\end{equation} 
For the physical case of $d=3$ and for a 4-particle processes (which should
dominate at late times in the models we have considered numerically, see
below) we obtain
\begin{equation}\label{p-EnCons-rth4}
p_{i} = \frac{1}{7} \; .
\end{equation} 
Note that for $d=2$ and $m=6$ we have $p_i = 1/11$, in agreement with
Eq.~\eqref{p-Mkindep-rel}.  For the exponent $s$ of particle distribution
functions in the energy~ cascade we find, see Eqs. (\ref{t-constE})
\begin{equation}\label{s-4-particle_En}
s = \frac{d+2}{m-1} = \frac{5}{3} \; .
\end{equation}

\subsection{Explicit time dependence in the collision integral}

The self-similar evolution is modified when an explicit time dependence is
present.  Below we consider two specific models with explicit time dependence
in the collision integrals which appear in the problem of reheating. The first
one is directly related to the relativistic scalar model we have simulated on
the lattice and time-dependence enters via the coupling to the zero-mode. The
second describes thermalization of non-relativistic particles and the time
dependence is induced by the expansion of the Universe.

\subsubsection{Non-zero classical field}
\label{sec:zero-mode}

Typically, oscillations of the inflaton zero-mode do not decay completely
during the initial stage of parametric resonance. Moreover, if the resonance
parameter is large, parametric decay stops early, when only a small part of
the initial inflaton energy has been transferred to particles
\cite{Khlebnikov:1997zt}. The remaining oscillating zero-mode serves as a
source in our turbulent problem. This source acts via two different
channels. The first one can be described as a direct decay into the resonance
band(s). The other channel is m-particle scattering when one or more particles
have zero momentum. These particles belong to the zero-mode (which is a Bose
condensate). While the zero-mode and excitations with $\bm{k} \neq 0$ can be
viewed as the same particles but with different momentum, the formal
description is different.  The presence of the zero mode $\phi_0$ leads to new
specific terms in the collision integral with reduced number of particles
participating in the interaction process and different (and time-dependent)
couplings.

The simplest example is 2 by 2 scattering in the $\lambda \phi^4$ model when
one of the incoming or outcoming particles belongs to the condensate.  These
scattering processes can be modeled as an effective 3-particle
interaction. The corresponding 3-particle collision integral can be obtained
from the 4-particle one with the substitution
\begin{equation}\label{nc}
\frac{n_{p}}{\omega_{p}} \rightarrow \,
\frac{n_{p}}{\omega_{p}}+(2\pi)^{3}\delta^{(3)}(\vec{p})\bar\phi_{0}^{2}\; .
\end{equation}
This gives an explicit time dependence in front of the collision integral, $B
= \phi_0^{2}(\tau)/\phi_0^{2}(1)$, and reduces the number of integrations by
one, $m=3$. Alternatively, the 3-particle collision integral in the background
of a zero mode can be derived from  first principles, see Appendix
\ref{WavesKinetics}.

The turbulent exponents for the 3-particle scattering without explicit time
dependence (i.e. $ \phi_0^{2}(\tau)=1 $), are given by
Eqs. \eqref{p-EnCons-3part} and \eqref{s-EnCons-3part}.  Both agree with what
is observed in our numerical experiments, see Sect.~\ref{NumSim:One-Field}. We
show in Sect.~\ref{CompLatt2Kin} that the collision integral in our lattice
problem is dominated by 3-particle interactions. Therfore, Eq.
\eqref{s-EnCons-3part} for the exponent $s$ seems to be indeed applicable for
the system considered numerically.  The question of applicability of
Eq. \eqref{p-EnCons-3part} for the exponent $p$ deserves special consideration
because the amplitude of the zero-mode changes with time.

During the initial stage, when the total energy in particles is small compared
to the energy stored in the zero-mode, we can consider the amplitude of
oscillations to be constant and the source of turbulence to be
stationary. However, distribution functions should then evolve with $p_t = 2
p_i$, see Eq. (\ref{p-turb2}).  At late times on the other hand we cannot
neglect the decay of the zero mode.  Numerical integrations show that the
amplitude of the zero-mode decreases as a power law, $\phi_0^{2}(\tau) \propto
\tau^{-\kappa}$. At late times this gives $p \rightarrow (1-\kappa) p$, see
Eq.~(\ref{A-GenSol1}). Numerically $\kappa = 2/3$, however, the conclusion
that $p = 1/15$ would be incorrect. First, for completely decayed zero mode
the 4-particle collision would dominate, leading to $p=1/7$. Therefore, in our
problem we should expect $p \geq 1/7$ at all times. Second, the condition $\tau
\gg 1$ is not fulfilled during our integration time. Indeed, we observed
self-similarity for $3600 < \eta < 10000$, see Fig.~\ref{self_sim}, which
corresponds to $\tau < 3$. For $\tau \approx 1$ the solution of
Eq.~\eqref{A-gSol}, \eqref{A-GenSol} for $A(\tau)$ coincides with $A =
\tau^{-p}$, while at $\tau = 3$ it deviates by not more than $5\%$.
Therefore, in this time interval $A(\tau) \approx \tau^{-1/5}$.  Similarly, the
quantity $ A^{\gamma} $ with $\gamma=4 $ for $1<\tau<3$ (energy conservation)
is close numerically to $\tau^{-q} $, where $ q\simeq 3/5$. Hence the indices
of self-similar evolution obtained in Sec.~\ref{NumSim:One-Field} are
explained by free turbulence driven by three-particle interactions in the
background of zero-mode.

\subsubsection{Non-relativistic regime in expanding universe}
\label{sec:NR_expanding} 
Let us consider now non-relativistic particles in an expanding universe with
physical dimension $d=3$.  We will be working in the conformal reference
frame, Eq. (\ref{metric}). In these coordinates the expansion of the universe
is simply accounted for by multiplying all bare mass parameters, $M$, by the
scale factor. This is true both for the original field equations and for the
kinetic equations (which are derived from the former). Factors of $\omega$ in
the measure Eq.~(\ref{U-def}) should be replaced by $M a(\eta)$. Therefore, in
the non-relativistic regime the collision integral in the expanding universe
can be obtained by multiplying it by the scale factor in some negative power.

In conformal reference frame the solution of the Friedmann equations for the
scale factor as a function of $\tau \equiv \eta/\eta_0$ can be written as
\begin{equation}\label{a-cf}
a^{b} = b H_0 \eta_0 (\tau - 1) + 1 \; ,  
\end{equation} 
where $H_0$ is the value of the Hubble parameter at time $\eta_0$.  For the
radiation dominated expansion $b=1$, while $b = 1/2$ for the matter dominated
expansion. Hence, the function $B(\tau )$ takes the form
\begin{equation}\label{B-friedmann}
B(\tau ) = [b H_0 \eta_0 (\tau - 1) + 1]^{-\kappa} \; .
\end{equation}
where $\kappa = 3/b$ for the 4-particle process in $\lambda \phi^4$
theory, i.e.  $\kappa = 3$ and $\kappa = 6$ for radiation and matter
dominated expansion respectively. This gives
\begin{equation}\label{intB}
\int_1^\tau B(\tau')d\tau' = 
\frac{1 - (b H_0 \eta_0 (\tau - 1) + 1)^{1-\kappa}}{b (\kappa -1) H_0 \eta_0}
\; .
\end{equation}
We see that in the limit $\tau \rightarrow \infty$
\begin{equation}\label{infA}
A(\tau = \infty ) = 
\left[ 1 + \frac{1}{b (\kappa -1) H_0 \eta_0} \right]^{-p} \; ,
\end{equation}
where $p$ is given by Eq.~(\ref{p-EnCons-non-rel}). The particle distributions
cannot propagate to high momenta and are frozen out at
\begin{equation}\label{infKc}
k_c(\tau =\infty) = \frac{k_c(1)}{A(\tau = \infty)} =
\frac{k_c(1)}{[b (\kappa -1) H_0 \eta_0 ]^{p}} \; .
\end{equation}

In the traditional discussion of thermalization of particles in the expanding
Universe, see e.g. \cite{kolbbook}, the expansion rate, $H_0$, is compared to
the to the rate of interactions, which in our case can be identified with
$\eta_0$ (see the normalization factor in Eq. \eqref{A-gSol}).  It is
concluded that particles can not thermalize if $H_0 \eta_0 > 1$ while they can
reach thermal equilibrium when $H_0 \eta_0 < 1$.  Equation (\ref{infKc}) tells
us that thermalization is indeed impossible for $H_0 \eta_0 > 1$ since the
distributions do not move towards high momenta in this case. However, it is
not guaranteed that the equilibrium is reached even if $H_0 \eta_0 \ll 1$. The
system may thermalize only if $k_c(\tau =\infty)$ is not smaller than the
typical values of momenta in eventual thermal equilibrium.


\section{Two Interacting Scalar Fields. Numerical Results.}
\label{TwoFields}

In this Section we present the results of lattice calculations of reheating in
the model of two interacting fields. As in the one field model presented in
Section~\ref{NumSim:One-Field}, we again consider the massless case, for which
the use of conformal transformation allows mapping of the dynamics in
expanding Friedmann universe into the case of Minkowski space-time. This
permits a long integration time on a fixed lattice.

\subsection{The model}
\label{TheModel}

At the end of inflation the universe is very close to a spatially flat
Friedmann model. It is convenient to work in conformal coordinates where the
metric takes the form $ds^2 = a(\eta)^2\, (d\eta^2 - dx^2)$.  We consider two
scalar fields $\Phi$ and $X$ whose dynamics are determined by the action
$\mathcal{S}=\int dt\, d^3x\, \sqrt{ -\mathit{ g } } \mathcal{L}(\Phi,X)$
with Lagrangian density
\begin{equation}
\mathcal{L}=\frac{1}{2} {g}^{\mu\nu}\partial_{\mu}\Phi
\partial_{\nu}\Phi\ + \frac{1}{2}
{g}^{\mu\nu}\partial_{\mu} X \partial_{\nu} X\ - 
{V}(\Phi,X)
\label{L}
\end{equation}
and potential
\begin{equation}
{V}(\Phi,X)= \frac{\lambda_{\Phi}}{4} \Phi^4 +  
\frac{\lambda_{\Phi X}}{2} \Phi^2 X^2 + \frac{\lambda_{X}}{4} 
X^4.
\label{pot}
\end{equation}

We identify the field $\Phi$ with the inflaton. Therefore $\lambda_{\Phi}\simeq
10^{-13}$ \cite{lindebook,kolbbook,lythbook}.  Inflation ends at time
$\eta_0$ when $\langle\Phi(\eta_0)\rangle\simeq 0.35 \, \mathrm{M_{{Pl}}}$.

We use the following set of coordinate and field rescalings
which bring the system into a dimensionless form suitable
for numerical integration:
\begin{equation}\label{Weyl-rescaling1}
\left.\begin{aligned}\,dx_0\;\\d{\bf x}_{i}\;\end{aligned} 
\right\} \;\longrightarrow\;  \left\{  
\begin{aligned}  
\;d\eta &\equiv dx_0\, \lambda_{\Phi}^{1/2}\Lambda \\
\;d{\bf y}_{i} &\equiv d{\bf x}_{i}\, \lambda_{\Phi}^{1/2}\Lambda \\ 
\end{aligned}\right.
\end{equation}
\begin{equation}\label{Weyl-rescaling2}
\left.\begin{aligned}\,\Phi\; \\ X\; \end{aligned} 
\right\} \;\longrightarrow\;  \left\{  
\begin{aligned}  
\varphi &\equiv \Phi\, \Lambda^{-1} a(\eta ) \\ 
\chi &\equiv X\, \Lambda^{-1} a(\eta )
\end{aligned}\right. 
\end{equation}
Re-scaling of the fields with $a(\eta )$ in Eq. \eqref{Weyl-rescaling2} rotates
the scale factor away and maps the model into a scalar field theory in
Minkowski space-time. The classical equation of motion have two independent
parameters
\begin{equation} 
{g} \equiv \lambda_{\Phi X}/\lambda_{\Phi}, {\rm ~~~~~~} h \equiv
\lambda_{X}/\lambda_{\Phi} 
\end{equation} 
and simplify to
\begin{eqnarray} 
\square \;\varphi + \varphi^{3}
+{g}\,\chi^2 \,\varphi &=&0\label{eq_mot_M:phi}  \; ,\\
\square \;\chi + h\,\chi^{3}
+{g}\,\varphi^2 \,\chi &=&0\; ,\label{eq_mot_M:chi} 
\end{eqnarray}
We choose $\Lambda=\langle\Phi(\eta_{0})\rangle$,
so that the initial condition for the inflaton zero-mode reads $\langle
\varphi(\eta_0 )\rangle=1$.  The equations \eqref{eq_mot_M:phi} and
\eqref{eq_mot_M:chi}, however, are independent on the particular choice of
$\Lambda$.  At $\eta = \eta_0$ all correlation functions of $\Phi$ and $X$ on
subhorizon scales characterize a vacuum of fluctuations around the inflaton
mean value.

\subsection{Results of numerical integration}
\label{NumSim:Two-Fields}

We have studied the two-field model using the following set of coupling
constants: $\lambda_{\Phi}=10^{-13}$, $g=30$, and $h$ was varied in the range
$0.1 g\leq h \leq 10^{4}g$.  We will see below that different values of $h$
lead to different duration and different relative importance of the specific
dynamical regimes, as it was already argued for in Sec.~\ref{Preview:other
models}. These are: the regime of parametric resonance, the regime of
stationary (or driven) turbulence and the regime of free turbulence.
These issues will be addressed later in this Section, which we start with the
discussion of particle spectra.

\subsubsection{Spectra}

\begin{figure}[t]
\includegraphics[width=3.9in]{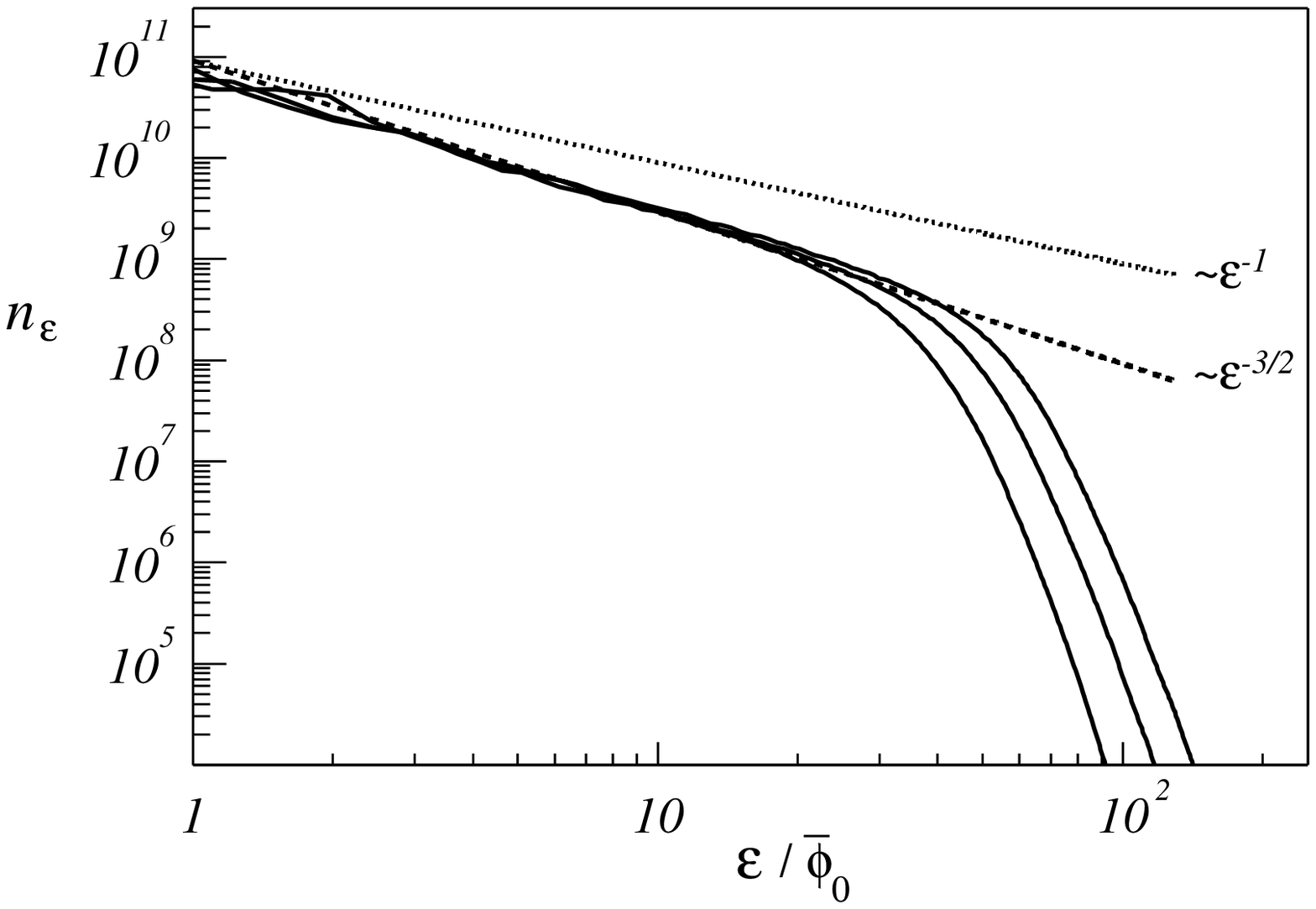}
\includegraphics[width=3.9in]{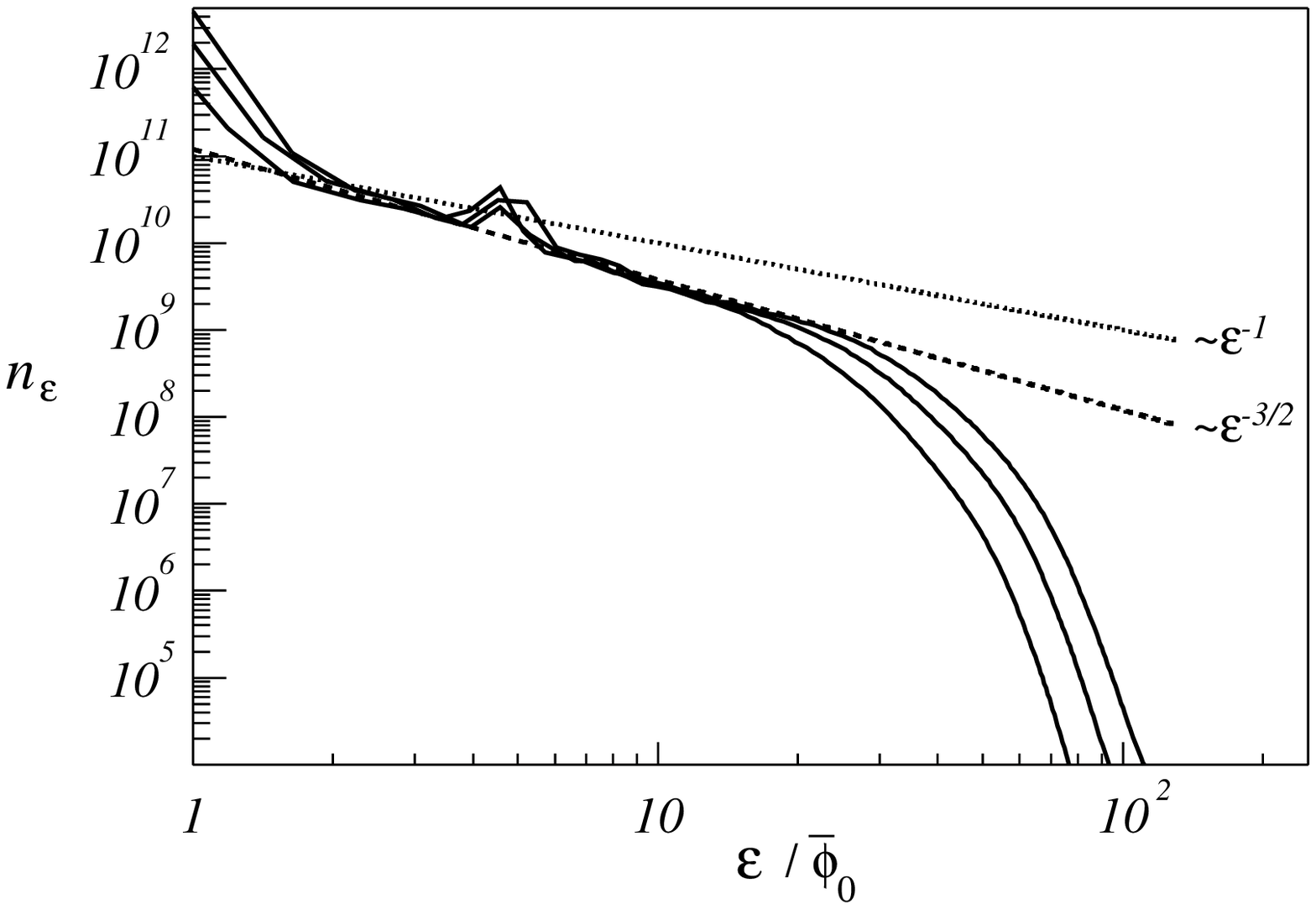}
\caption{Particle distributions in the self-similar regime for $h=10g$ as
functions of the corresponding wave kinetic energies rescaled by the current
zero-mode amplitude $\bar\phi_{0}$.. Upper and lower panels correspond to
$\chi$ and $\phi$ fields respectively.  In both cases from left to right the
plots are taken at $\eta=1000,1500,2000$.  }\label{fig:spectra}
\end{figure}
\begin{figure}[t]
\includegraphics[width=3.9in]{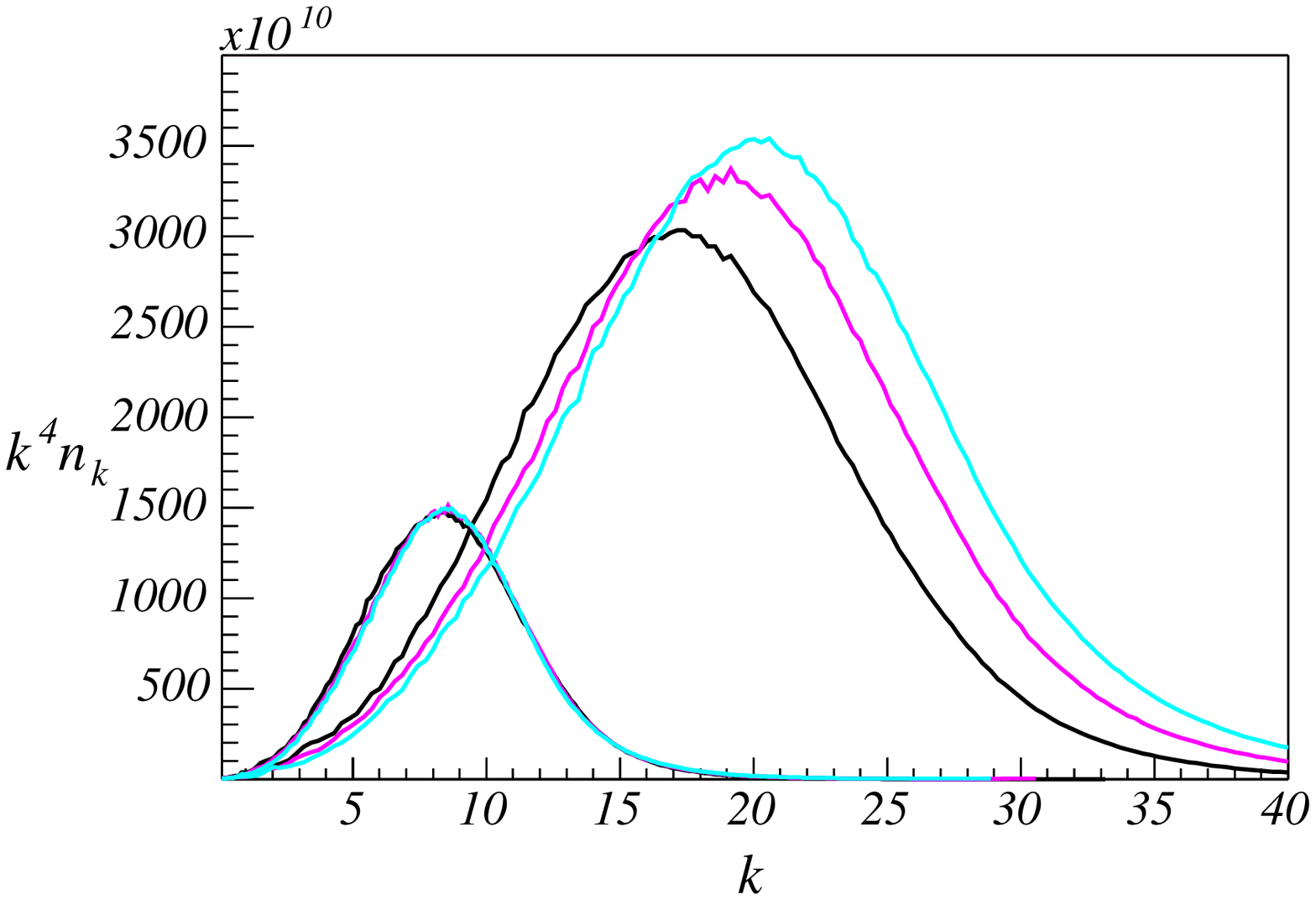} 
\includegraphics[width=3.9in]{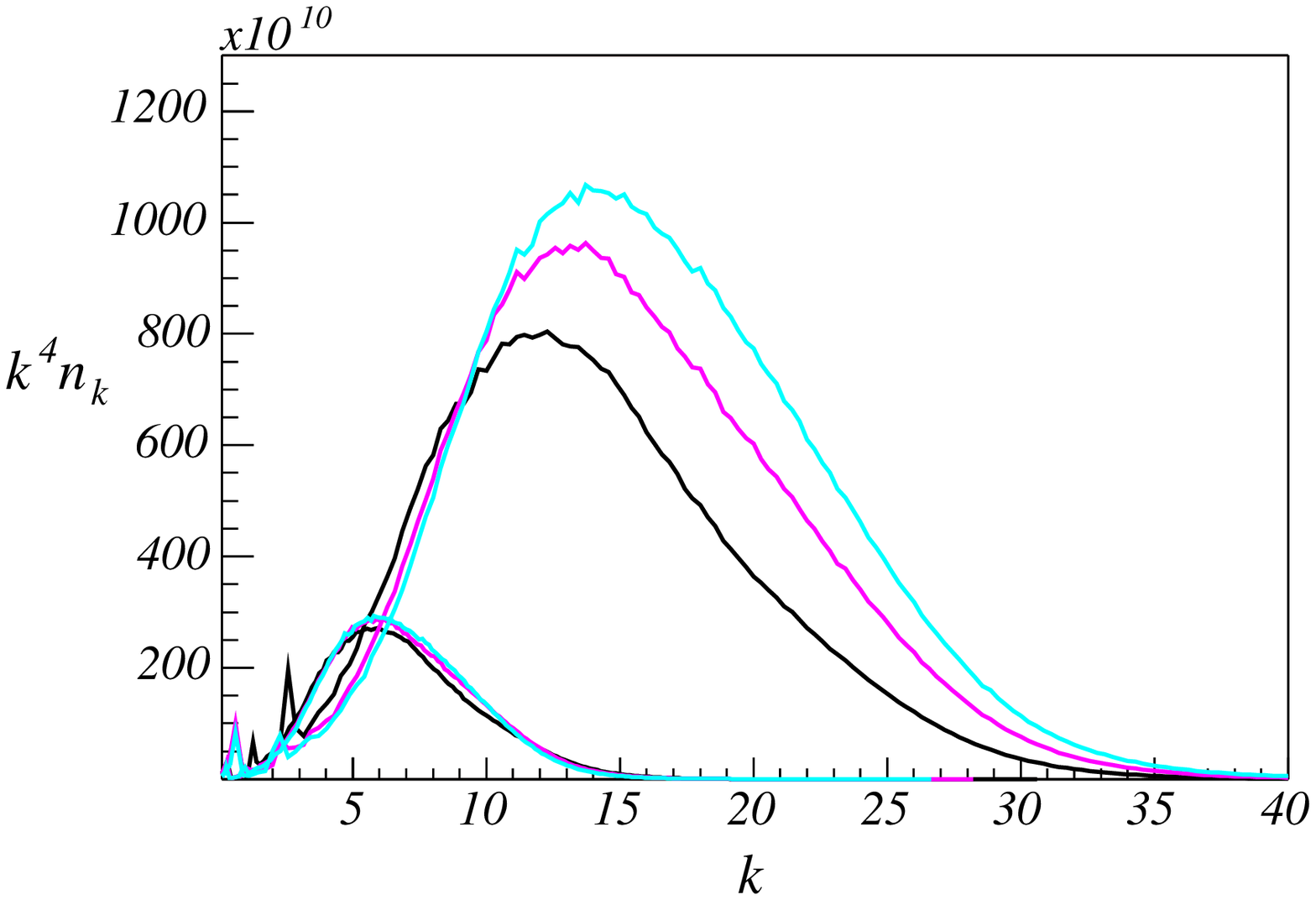}
\caption{Spectral energy distributions for $\chi$ (upper panel) and $\phi$
(lower panel) in the model with $h = 10 g$.  In each panel we plot the wave
energy per decade found in lattice integrations at three moments of time,
$\eta=1000$, $1500$ and $2000$. In the lower-left corner of each panel are the
same graphs transformed according to the relation inverse to
Eq.~(\ref{SelfS}). }\label{fig:selfsim}
\end{figure}

The particle spectra in the two field model at late times are very similar to
what we have observed in the one field model and have the same turbulent
exponents.  Namely, in the inertial range $n_k$ is a power low with the
exponent $s = 3/2$, for both fields $\chi$ and $\phi$, see
Fig.~\ref{fig:spectra}. And both fields evolve in a self-similar way with $p =
1/5$ at sufficiently late times, when the energy in particles became
comparable to the energy in the zero-mode, see Fig.~\ref{fig:selfsim}. Both
exponents, $s$ and $p$, correspond to turbulence supported by 3-particle
interactions.

There are some differences however. For the considered range of parameters,
the coupling of the excitations to the medium is rather strong, which induces
large effective particle masses, see Appendix \ref{ContStochInCon}. Therefore
particles are non-relativistic already in the part of the inertial
range. Namely, $M_{\chi}\simeq 5.5\bar\phi $ and $M_{\phi}\simeq 1.7\bar\phi
$.  This manifests itself as $\sim k^{-3}$-power-law behavior, which is again
consistent with domination of 3-particle interactions, see
Eq. \eqref{p-Mkindep-nonrel}.  This can be expressed as a single power law if
particle distributions are plotted as functions of relativistic kinetic
energy,
\begin{equation}\label{KinEnDef}
\epsilon_k \equiv \omega_k - M \; ,  
\end{equation}
where $M$ is the effective particle mass.  Indeed, in the relativistic region
we have $n_k \propto k^{3/2} \propto \epsilon_k^{3/2}$, while in the
non-relativistic region we obtain $n_k \propto k^{3} \propto
\epsilon_k^{3/2}$. For this reason, the particle distributions were plotted in
Fig.~\ref{fig:spectra} as functions of $\epsilon_k$. The particle
distributions for the ${\chi}$ field appear in this variable as featureless
single power law. This can be easily understood. First, the energy transport
for 3-particle interactions in the presence of zero mode corresponds to the
transport of kinetic energy, as energy conservation law in elementary
scattering process, which involves the frequency of zero-mode oscillations,
$\omega_0 \approx M$, tells us, see Appendix~\ref{WavesKinetics}. Second, the
collision integral, Eq. \eqref{I3.0}, being substituted into expression for
the energy flux, Eq. \eqref{Frho_def}, will have appropriate universal scaling
behavior in terms of kinetic energy, $\epsilon_k$, but not in terms of
$k$. Therefore, the kinetic energy is indeed the appropriate variable for the
case of 3-particle interactions in the presence of zero mode.

For $h > g$ the spectra look stationary in the inertial range after rescaling
$\epsilon$ by the current zero-mode amplitude $ \bar\phi_0\sim \eta^{-1/3}
$. This is similar to the one field case, (see fig.~\ref{fig:spectra}).
However, for $h\leq g$ we found $\bar\phi_{0}\sim \eta^{-2/3}$, but the
spectra still appear stationary after rescaling by $\eta^{-1/3}$.  This can be
understood in the light of Eq.~\eqref{renorm-spectra}: $b(\tau) = \tau^{-1/3}$
is consistent with the choice $\gamma=4$, $s=3/2$ and $p= 1/5$. Hence, the
decreasing amplitude of distribution functions in the region of low $k$ simply
reflect the energy conservation in the system.

\subsubsection{Stationary and free turbulence regimes}

Let us demonstrate now that the regime of stationary turbulence does occur in
the two field model. This regime is expected to appear in the case of large
values of dimensionless parameters, $g \gg 1$, $h \gg 1$, when parametric
resonance stops early, while the total energy is still stored in the zero
mode.

We found that in the relevant range of parameters the description in terms of
particles, which we were using so far, deteriorates. The reason is that in
this language at large couplings there is no unique way to split the total
energy density of the system into contributions coming from zero mode and
fluctuation field.

\begin{figure}[t]
\includegraphics[width=3.9in]{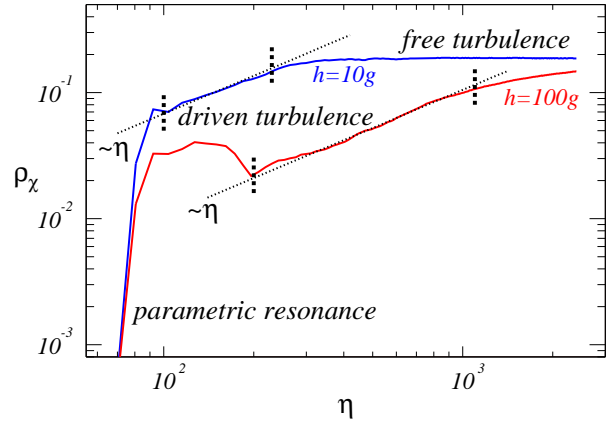}
\caption{Different regimes of the evolution of the $\chi$ field for two values
  of self-coupling, $h=10g$ and $h=100g$. The dashed lines correspond to a
  linear growth of energy in the $\chi$ field with time, $\rho_\chi \propto
  \eta$.  }\label{fig:energy}
\end{figure}
\begin{figure}[t]
\includegraphics[width=3.9in]{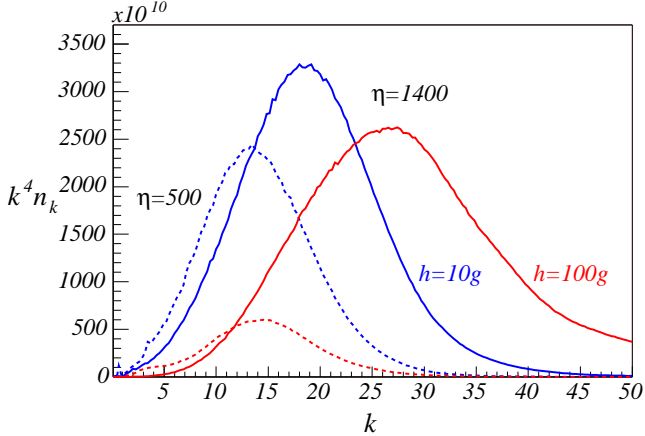}
\caption{Spectral energy distributions at two moments of time,
$\eta = 500$ (dotted lines) and $\eta = 1400$ (solid lines). We compare two
models with different self-coupling, $h=10g$ and $h=100g$. 
}\label{fig:SED-two-h}
\end{figure}

To deal with this problem we have quantified the energy transfer in the
following way. The quantity $\rho_{0} \equiv \frac{1}{2}{\dot\phi_0
(\eta_z)}^{2}$ gives a good measure of the total energy density stored in
zero-mode oscillations, if it is measured at those moments of time,
$\eta_{z}$, when the mean field crosses zero, $\phi_0(\eta_{z})=0$.  In this
way we can get rid of the ambiguity in accounting interaction energy between
zero-mode and fluctuations. Similarly, we measure the energy density in the
fluctuation field as $\rho_{\chi}\equiv\langle{\dot\chi}^{2}\rangle_{t}$ and
$\rho_{\phi} \equiv \langle{\dot\phi}^{2} \rangle_{t}$ for the $\chi$ and
$\phi$ fields respectively. Here $\langle\ldots\rangle_t$ means lattice and
time averaging.  We verified numerically that the sum of these quantities
conserves with time and equals to the initial energy density.  This is not
true, however, when we measure the energy density in particles as $\omega_k
n_k$. Both measures of particle energy converge at late times when the
interaction energy becomes unimportant.

This "kinetic" measure of the total energy density stored in particles as a
function of time is shown in Fig.~\ref{fig:energy}.  We compare models with
two different values of $h$. Three different regimes are clearly seen in
both cases.

\begin{enumerate}
\item \textit{Parametric Resonance:} The energy density $\rho_{\chi}$ grows
exponentially. This regime continues until re-scattering becomes important.
The larger $h$ is, the earlier resonance terminates.

\item \textit{Stationary Turbulence:} At later time the energy density in
${\chi}$ particles grows linearly in time, which according to
Eq.~\eqref{Epropt} is a sign of stationary turbulence. During this period the
energy density still stored in the zero-mode dominates the total energy
balance.

\item \textit{Free Turbulence:} At some point the energy density in the
zero mode drops below the energy density already stored in
particles. Stationary turbulence cannot be sustained anymore and the regime of
free turbulence, with conserved energy in particles, follows.  We may
expect self-similar evolution of particle distribution functions, which at
late times are good quantities.

\end{enumerate}  

In the model with larger self-coupling the parametric resonance stops earlier
and only a negligible part of the inflaton energy is transferred to particles
during the resonance stage, see Fig~\ref{fig:energy}. In this parameter range
the transfer of energy from the inflaton into $\chi$-field is dominated by a
stationary turbulence. In the Sec.~\ref{Thermalization} we show that if all
coupling constants are of order of the inflaton self-coupling, the
thermalization is a very long process and the Universe reheats to unacceptably
low temperature, $T\sim 100$ eV. Therefore, some couplings in the sector of
physical fields (e.g. self-couplings, or couplings to the inflaton) in a
realistic model have to exceed significantly the scale of the inflaton
self-coupling. With larger couplings the thermalization proceeds faster.  This
is confirmed in our lattice integration, see Fig.~\ref{fig:SED-two-h}. At
earlier times the model with larger self-coupling contains less energy in
$\chi$-particles, cf. curves at $\eta = 500$. However, at later times this
model takes over and the energy containing region moves faster towards 
ultraviolet in the model with larger self-coupling.

With even larger self-coupling of the $\chi$-field, or its coupling to the
inflaton, the period of stationary turbulence should become even more
pronounced. In light of these findings, we can also understand the results of
earlier papers \cite{Khlebnikov:1997zt,Prokopec:1997rr}. E.g., in Figures
presented in Ref.~ \cite{Prokopec:1997rr}, we see clear signs of driven
turbulence, which was not identified as such until now. In particular, in
Ref.~ \cite{Prokopec:1997rr} it was found that the energy in
$\chi$-fluctuations grows with time as $\rho_\chi \propto t^{0.95}$. (Small
deviation from $\propto t$ law can be due to the fact that the energy in zero
mode decreases somewhat and the source deviates from stationarity). This
regime persists until the final integration time, when distribution functions
reach the boundary of the integration box, and even then the system is far from
free turbulence regime.

We conclude that in the models with an acceptable reheating temperature, the
parametric resonance stops only when a negligible fraction of the inflaton
energy has decayed. Therefore, in  {realistic models} of the type considered
in the present paper, the {major} mechanism of {energy transfer} from the
inflaton into particles is {stationary turbulenc}e.

\section{Is the kinetic approach applicable ?}
\label{CompLatt2Kin}

In this section we confront the results of our lattice integration with the
predictions of kinetic theory and address the validity of the kinetic
description at the thermalization stage during our integration time interval.

The particle distributions in the inertial range, $n(k) \sim k^{-s}$ with $s
\approx 3/2$, which we observe in the lattice simulations, can be understood
as corresponding to the scale invariant energy flux for 3-particle
interactions, see Eq.~(\ref{s-EnCons-3part}). The observed exponent $p =1/5$
of the self-similar scaling of free turbulence, can also be in accord with
3-particle interactions, see Eq.~(\ref{p-EnCons-3part}). However, in our case
bare 3-particle couplings are absent and appear effectively in interactions
with zero-mode. Therefore, the 3-particle collision integral is multiplied by
the amplitude of zero mode squared. Since the amplitude of the zero mode
oscillations decays, one can expect $p = 1/5$ only during a small time
interval, see Sect.~ \ref{sec:ExplTimeDep}.

Can 4-particle interactions be responsible for the observed scalings ?  For
4-particle interactions $p_i =1/7$, see Eq. (\ref{p-EnCons-rth4}), which is
not that far away from the lattice results, especially if one takes into
account energy influx from the zero-mode. However, for particle distributions
in the inertial range one should expect $s = 5/3$, which is not in a good
agreement with the observed value of $s =3/2$.  Further, in view of
Eq.~(\ref{nc}) one should expect the dominance of 4-particle scattering during
the time interval when the variances of fluctuations are larger than
$\phi_0^{2}$. This is not the case during the time interval encompassed by the
lattice simulations, see Fig.~\ref{spat_av}.

The outlined difficulties may give an indication that the weak turbulence
description is not applicable in our case. In view of the importance of the
issue, we performed a detailed study of collision integrals, anomalous and
higher order correlators, as measurements on the lattice, and compared these
with predictions and assumptions of kinetic theory.

\subsection{Collision integrals}

To verify the extent of agreement between kinetic theory and lattice
calculations, and to find out which processes dominate the collision integral
in our problem, we carry out the following procedure. First, we numerically
calculate the collision integrals using standard expressions,
Eqs. (\ref{coll-int:def1})-(\ref{FLargeN}), and the particle distribution
functions $n_k (\eta )$ extracted from our lattice calculations. Second, using
lattice data we calculate time derivatives of the distribution functions to
see if the relation $\dot n_{k} = I_{k}[n]$ holds.  We limit ourselves to 3-
and 4-particle collisions.

The general relations, Eqs. (\ref{coll-int:def1})-(\ref{FLargeN}), for 4- and
3-particle collision integrals can be reduced to two and one dimensional
integrations respectively, if the distribution functions are
isotropic. Explicit expressions are given in Appendix~\ref{WavesKinetics} ,
Eqs.~\eqref{I3.0} and \eqref{I4.0}.

\begin{figure}
  \includegraphics[width=3.9in]{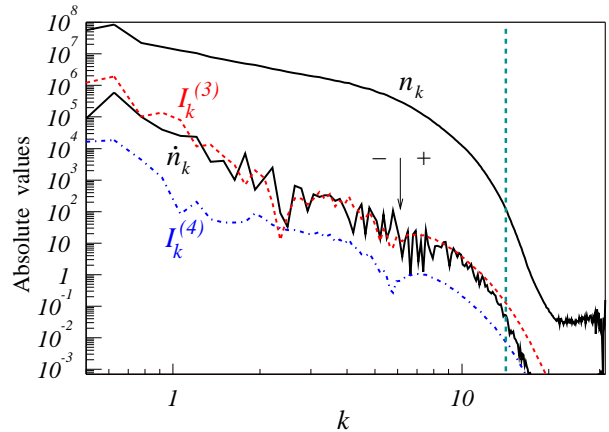}
  \caption{Absolute values of $\dot{n}(k)$ and of $I_k^{(3)}$ and $I_k^{(4)}$
collision integrals at $\eta = 5000$. To the left of the arrow $\dot{n}(k)$
and collision integrals are negative, while to the right they are positive.
Occupation numbers, $n_k$, are also shown for comparison. $I_k^{(3)}$ agrees
with $\dot{n}(k)$ to the left of the vertical dashed line.  }
\label{I3andI4}
\end{figure}

The numerically calculated values of $I_k^{(3)}$ and $I_k^{(4)}$ collision
integrals are shown in Fig.~\ref{I3andI4} in comparison with $\dot{n}_k$.
Note that the collision integrals and $\dot{n}_k$ take positive and negative
values. For clarity we show only absolute values of these functions and
indicate schematically the boundary between regions where $\dot{n}_k$ is
negative and positive. Roughly, in the inertial range $\dot{n}_k$ is negative
(recall that in this region the particle distributions can be approximated as
$n_k(\eta ) = (\bar{\phi}_0/k)^{s}$ and are decreasing functions of time),
while $\dot{n}_k$ should be positive at larger k where the cut-off starts
(recall that energy is flowing into this region).

We find that $I_k^{(3)}$ gives a reasonable approximation to $\dot{n}_k$
practically in all range of $k$ which is dynamically important, which is to
the left of the vertical dashed line in Fig.~\ref{I3andI4}.  One reason for
the disagreement between $\dot{n}(k)$ and $I_k^{(3)}$ at larger $k$ could be
due to the fact that on the lattice some of the allowed resonant wave
interactions of the continuum limit are not present ( cf.~\cite{Pushk:2000}).
In any case, in the region where $I_k^{(3)}$ and $\dot{n}_k$ disagree, the
occupation numbers are relatively small, $n_k < 10^2$, and this region should
not contribute to the dynamics significantly.

The $I_k^{(4)}$ collision integral is about an order of magnitude smaller
compared with $I_k^{(3)}$ and is subdominant in the evolution of $n_k$, except
on the very tail of the distribution, see Fig.~\ref{I3andI4}.  The agreement
between $\dot{n}_k$ and $I_k^{(4)}$ in the region of the tail is not
coincidental - we observe it at all $\eta$.

\subsection{Anomalous correlators}
\label{anomalouscorrelators} 

\begin{figure}
  \includegraphics[width=3.9in]{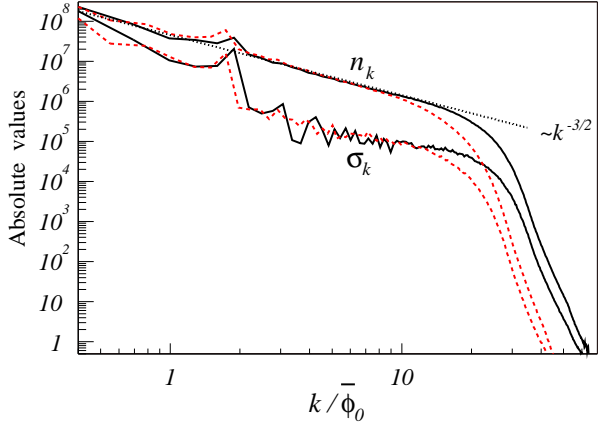}
  \caption{Occupation numbers and absolute values of $\sigma_k$
    at $\eta = 5000$ (dashed lines) and $\eta = 10000$ (solid
    lines). }
\label{fig:sigma}
\end{figure}

Usually, kinetic equations are derived under the assumption $\langle a_k
a_q \rangle \ll \langle a_k^{*}
a_q \rangle$. However, this condition not always holds. For example, in
the case of particle creation by a time-varying classical background (e.g. in
the region of parametric resonance)
\begin{equation}\label{ndot-AnomCorr}
\dot{n}_k = \frac{\dot{\omega}_k}{{\omega_k}}\, {\rm Re} 
(\sigma_k )\, ,
\end{equation}
where
\begin{equation}\label{AnomCorr}
\langle a_k a_q \rangle \equiv \sigma_k \, \delta ({\bf k}+{\bf q}) \, ,
\end{equation}
see Appendix~\ref{WavesKinetics}. In this case the anomalous correlators,
$\sigma_k$, can not be neglected, since $\sigma_{k} \sim n_{k}$. This holds in
general: if coherent processes are important, the correlators
Eq. (\ref{AnomCorr}) may modify the dynamics of $n_k$. If this is the case,
they should be included into the kinetic equation.  Since $\sigma_k$ were
neglected in the kinetic equations,
Eqs. (\ref{coll-int:def1})-(\ref{FLargeN}), it is important to verify if the
condition $|\sigma_k| \ll n_k$ holds in our simulations.

The correlators $\sigma_k$ are shown for several moments of time in
Fig.~\ref{fig:sigma}. In the inertial range the anomalous correlators are
small indeed, $|\sigma_k|/n_k \approx 3\cdot 10^{-2} $, while this ratio is an
order of magnitude larger in the region of the resonance peak ($k \approx 0.5$
at late times), which is expected behavior.  The ratio $|\sigma_k|/n_k$ is
growing also in the region of large $k$, reaching the value of 0.1 at $k = 8$
at late times, see Fig.~\ref{self_sim}. To avoid confusion, note that $k = 8$
corresponds to $k/\bar{\phi}_0 \approx 25$, which is the variable used in
Fig.~\ref{fig:sigma}. We do not know if the growth of $|\sigma_k|/n_k$ at
large $k$ is a lattice effect, but we can conclude that the kinetic equations
in its simple form, Eqs. (\ref{coll-int:def1})-(\ref{FLargeN}), should be
applicable in the inertial range.

\section{Physical Applications}
\label{PhysApplications}

Many different effects may occur during the stage of preheating.  Some of
these were discussed in the Introduction section. They have a common physical
origin: rapid particle creation and large accompanying fluctuations of the
classical fields involved. These findings are unaffected by our results, even
in the case when only a relatively small fraction of the inflaton energy is
transferred to  fluctuations during the initial stage of parametric
resonance.

However, in many cases it is necessary to trace the events further in time,
e.g. to find out when and how the symmetry which was restored during
preheating gets broken later on, or to trace which fraction of baryon or
lepton asymmetry survives in the process of thermalization. Finally, one needs
to know when thermal equilibrium will be established. This gives e.g. the
abundances of particular dark matter particle candidates and other, possibly
cosmologically ``dangerous'' relics like the gravitino.

The explicit time dependence of the particle distribution functions, and the
knowledge that the evolution is self-similar, $n(k,\tau) = A^{\gamma}\,n_0(kA)
= \tau^{-\gamma p}\,n_0(k \tau^{-p})$, which we have found in the present
paper, may be useful here. Below we discuss some applications, limiting
ourselves to field variances and to the problem of thermalization.

\subsection{Field variances}
\label{sec:Vars}

In some applications, basic observables like field variances may already give
the answer to the problem in question. This applies to the problem of symmetry
restoration.  To illustrate this, let us consider the Higgs field which is
coupled to a $\chi$-field. In the vacuum state without condensate the mass
squared of the Higgs field would  be negative, $-\mu^2$, and the
corresponding symmetry is broken. In the presence of the background of
$\chi$-particles, the mass gets ``dressed'', $m_{\rm eff}^2 = -\mu^2 + g
\langle \chi^2\rangle$. If the field variances are sufficiently large, the
symmetry is restored (and is broken when $\langle \chi^2\rangle \le
\mu^2/g$). 

If anomalous correlators are negligible, the field variances can be
calculated using expression
\begin{equation} 
var(\chi) ~\equiv~ \langle\chi^2\rangle - \langle\chi\rangle^2 ~=~ \int
  \frac{d^d k}{(2\pi)^{d}}\, \frac{n_k}{\omega_k} \; .
\label{variance-def}
\end{equation}
With the help of the self-similar substitution, Eq.~\eqref{ss1}, we find
\begin{equation} 
var(\chi,\tau)  =  A^{\gamma -d + \alpha}\,  var_{0}(\chi) \; .
\label{variance-ss1}
\end{equation}
Here, the left hand side is taken at conformal time $\eta $, while $
var_{0}(\chi) $ on the right hand side is the variance at some earlier time
$\eta_{0} $.

\subsubsection{Relativistic regime}

\begin{figure}
  \includegraphics[width=3.9in]{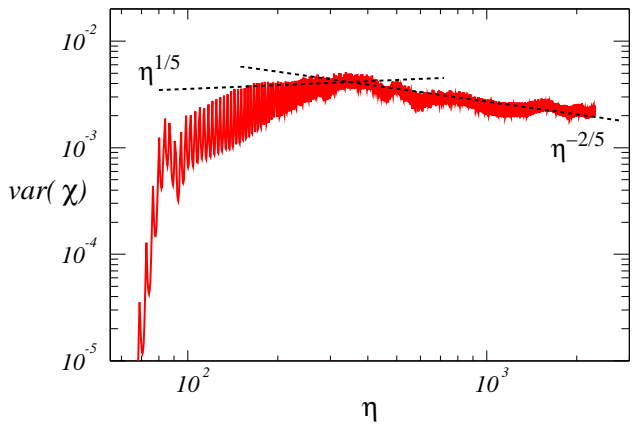}
  \caption{Time dependence of the variance of $\chi$ field  in 
the model with $h = 10 g$ considered in Sec.~\ref{TwoFields}. }
\label{fig:variance-2field}
\end{figure}

\paragraph{Free turbulence.} 
In this case $\gamma = d+\alpha$, see Eq.~\eqref{gamma-EnCons}, and we find
with $\alpha=1$
\begin{equation} 
var(\chi,\tau) = A^{2\alpha}\, var_0(\chi) = \tau^{-2p_{i}}\, var_0(\chi)\; .
\label{variance-ss2}
\end{equation}
For systems that we have studied numerically, $p_i = 1/5$ at early times which
span the integration period. Therefore, in the free turbulence regime, we
should expect $ var(\chi ,\tau)= \tau^{-2/5}\, var_0(\chi)$.  This is in
agreement with the results of our numerical integration, see
Fig.~\ref{spat_av}.  For late-time evolution, when 4-particle interactions
will start to dominate, we predict a slower decrease of the variance, $ var(\chi
,\tau)= \tau^{-2/7}\, var_0(\chi)$.

Note that these results have to be divided by the scale factor squared if
expansion of the Universe is important.

\paragraph{Driven turbulence.} In the regime of stationary turbulence 
without sink, according to Eq.~\eqref{gamma-stationary}, $\gamma = s$, which
gives $var(\chi,\tau) = A^{s-d+\alpha}\, var_0(\chi)$.  Using
Eq.~\eqref{s-Mkindep-rel}, we find $s-d+\alpha = 1/(1-m)$ and
\begin{equation} 
var(\chi,\tau) = \tau^{-p_t/(1-m)}\, var_0(\chi) = \tau^{p_i} var_0(\chi)\; ,
\label{variance-driven}
\end{equation}
where relation Eq.~\eqref{p-turb2} was also used.  Therefore, during driven
turbulence the variance should grow as $ var(\chi,\tau) = \tau^{1/5}\,
var_0(\chi)$ in the models which we have integrated numerically. This is indeed
the case as Fig.~~\ref{fig:variance-2field} shows. The transitional period
from the regime of parametric resonance to the regime of stationary turbulence
at $\eta \sim 10^2$ is slightly more pronounced in
Fig.~~\ref{fig:variance-2field} as compared to Fig.~~\ref{fig:energy}. This
may be explained by the fact that different regions of momentum space are
emphasized in $\rho_\chi$ and in $var(\chi)$.

\subsubsection{Non-relativistic regime}

In the case of free turbulence we have $\gamma = d$, and in
Eq.~\eqref{variance-ss1} we have to substitute $\alpha = 0$, which corresponds
to $\omega_k \rightarrow M$ in Eqs.~\eqref{EnCons}, \eqref{variance-def}.
Therefore, $var(\chi, \tau) = {\rm const}$.  For driven turbulence we have
$\gamma = s = d$, see Eq.~\eqref{s-Mkindep-nonrel}, and
Eq.~\eqref{variance-ss1} again gives $var(\chi, \tau) = {\rm const}$.

~\vspace{1pt}

We see that in the regime of driven turbulence variances are slowly changing
functions of time ($\propto \tau^{1/5}$ in the relativistic case and
$var(\chi, \tau) = {\rm const}$ in the non-relativistic case), while energy in
particles grows fast, $\rho_\chi \propto \tau$ in this regime. This is in
accord with the fact that variances can be large right after the initial
parametric resonance stage, while the amount of energy transfered during this
stage is low and all energy transfers occur in the regime of driven
turbulence.

\subsection{Thermalization in the absence of zero-mode}
\label{Thermalization}

We now apply the results obtained earlier in this paper to the general problem
of thermalization of relativistic and non-relativistic scalar particles, both
in Minkowski space-time and in expanding Friedmann universe. We do not
restrict ourselves to the models which were studied numerically. Our analysis
will be based on expression \eqref{ss1} with the factor $A(\tau )$ being
specified for a particular modeled. This expression describes a self-similar
propagation of the distribution functions into the ultra-violet. In a
classical theory this evolution continues without bound (unless we consider a
non-relativistic theory in expanding universe).

The classical evolution stops when a system reaches the quantum regime where
it can relax to the Bose-Einstein distribution.  We adopt that this happens
when in a region of momenta, $k_f$, which saturate the energy integral, the
occupation numbers became of order one.  In this subsection we consider the
case of free turbulence. Then, one can estimate $k_f$ using energy
conservation and approximating the energy density as $\rho \sim \omega(k_f)
k_f^3$ in the region where $n_k\sim 1$.  On the other hand, initially the
energy was deposited into particles with lower momenta, denoted below as
$k_i$. The relation $k_i \sim M_\phi$, where $M_\phi$ is the inflaton mass,
determines the scale of initial momenta.  Eq. \eqref{k_c-motion} gives for the
time needed to thermalize a system:
\begin{equation}
\tau^{\rm th} \sim (k_f/k_i)^{1/p}
\label{ThermalizationTime}
\end{equation}
Actually this should be considered as a lower limit on the thermalization time
since we have to add a time which the system will spend in the quantum regime.

As an idealization of the thermalization process we consider the evolution of
a sub-system of excitations of a field $\chi$, assuming that a fixed part of
energy was deposited into it initially, while since then $\chi$ evolves as an
isolated system. In this subsection, for estimates of the thermalization time
we neglect the presiding regimes of  parametric resonance and of 
stationary turbulence, since they are much shorter if the relevant coupling
constants are not drastically different. We consider the possibility of
(partial) thermalization in the regime of driven turbulence in the following
subsection.

As a first step we will find the thermalization time which follows from the
exact self-similar solutions obtained above. Then we will show that in all
cases which we consider, the result coincide, parametrically, with the
``naive'' perturbative estimates. Doing this comparison we neglect all
numerical coefficients.

\subsubsection{Relativistic regime}

Equation (\ref{ThermalizationTime}) gives
\begin{equation}
\tau^{\rm th} \sim (\rho_f^{1/4}/M_\phi)^{1/p} \;.
\label{ThermalizationTimeRel}
\end{equation}
The expansion of the Universe is easily treated in conformal reference
frame. We have $\rho_f = \rho_i = c_\chi \rho_{\rm tot}$, where $c_\chi$ is
the fraction of the inflaton energy deposited into the field $\chi$ during
preheating and driven turbulence.  This finalizes the answer. The result is
general and is valid for any model. The initial inflaton energy can be written
as $\rho_{\rm tot} \sim k_i^2 \phi_0^2 \sim M_\phi^2 M_{\rm Pl}^2$, where
$\phi_0$ is the initial amplitude of inflaton oscillations. We find with
$p=1/7$, Eq. \eqref{p-EnCons-rth4}, which corresponds to a relativistic theory
with dimensionless couplings
\begin{equation}\label{tauTerm}
\tau^{\rm th} \sim c_\chi^{7/4}\left(\frac{M_{\rm Pl}}{m_\phi} \right)^{7/2}
\sim c_\chi^{7/4} 10^{21}  \, . 
\end{equation}
We used here the inflaton parameters, $M_\phi \approx 10^{-6} M_{\rm Pl}$ in
$M_\phi^2 \phi^2$ model, or $M_\phi = \sqrt{\lambda} M_{\rm Pl}$ with $\lambda
\approx 10^{-13}$ in the $\lambda \phi^4$ inflationary model.

To avoid confusion, note that the definition of $\tau$ is different in
different models since it involves $t_0 \sim \Gamma^{-1}$.

We can assume $c_\chi \approx 1$ if the Universe expansion can be neglected
(this may be of interest for problems outside of inflationary cosmology), and
if the number of competing channels (other fields beyond $\chi$ to which the
initial energy can be deposited) is not large.

\paragraph{Minkowski space-time}

Let us show that Eq. (\ref{ThermalizationTimeRel}) agrees with the "naive''
perturbative estimate. For this estimate we define $\tau$ as $\tau =t_R\,
\Gamma$, where $t_R$ is the perturbative estimate of the thermalization time
$t_R^{-1} \sim \sigma n$. In the $\lambda \phi^4$ model $\sigma \sim
\lambda^2/T^2$, $n \sim T^3$, $T \sim k_f$ and therefore $t_R^{-1} \sim
\lambda^2 k_f$. On the other hand, parametric resonance stops when the rate of
re-scattering from the resonance band becomes equal to the rate of particle
production $\mu \sim M_\phi \sim k_i$.  This gives $\Gamma \sim k_i$ and we
find $\tau \sim \Gamma t_R \sim k_i/\lambda^2 k_f$.  Now, $\rho \sim k_f^4$
and $\rho \sim k_i^4 n_k$, where $n_k$ correspond to the typical occupation
numbers at the time when parametric resonance stops, $n_k\sim 1/\lambda$.  We
obtain $t_R \Gamma = \left( \rho^{1/4}/k_i \right)^7$, in agreement with
Eq. (\ref{ThermalizationTimeRel}).

\paragraph{Friedmann universe}

In this case we can estimate the final temperature as $T \sim k_f/a(\tau
)$. Let us consider a radiation dominated universe with $ a(\tau ) = H_0
\eta_0 (\tau -1) + 1$, see Eq. (\ref{a-cf}).  We neglect the rapid epoch of
stationary turbulence, and $\eta_0$ corresponds to a time when the evolution
of $\chi$ is driven by its self-interaction with self coupling $\lambda_\chi$,
i.e. $\eta_0^{-1} \sim \Gamma \sim \lambda_\chi^2 n_k^2 k_i^{~} \sim
\lambda_\chi^2 (c_\chi \rho^{~}_{\rm tot})^2 /k_i^7$, where we have used
$\rho_\chi \sim k_i^4 n_k$.  On the other hand $H_0 \sim \rho_{\rm tot}^{1/2}
/M_{\rm Pl}$.  Combining this with Eq. (\ref{ThermalizationTimeRel}) we find
$a(\tau ) = H_0 \eta_0 \tau \sim c_\chi^{-1/4} \rho_{\rm
tot}^{1/4}/\lambda_\chi^2 M_{\rm Pl}$. For the final thermalization
temperature we obtain $T \sim k_f/a(\tau ) \sim c_\chi^{1/2}\lambda_\chi^2
M_{\rm Pl}$, where we have used $k_f \sim \rho_\chi^{1/4}$.  This again agrees
with the naive estimate, $\sigma n \sim H$.

Numerically, $T \sim \lambda_\chi^2 M_{\rm Pl} \sim 100$ eV, if we use the
strength of the inflaton self-coupling, $\lambda \approx 10^{-13}$.  Therfore,
in a realistic model, at least some couplings should be significantly larger
than this scale. 

\subsubsection{Non-relativistic regime}

Now we consider the thermalization of X-particles of mass $M_X$ in the
non-relativistic regime. We assume that the relaxation is due to the
self-interaction $\lambda_X X^4$. The particle number conserves in the
conformal reference frame in this regime, and Eq. (\ref{ThermalizationTime})
gives
\begin{equation}
\tau^{\rm th} = c_X^{1/3p} 
\left[\frac{1}{M_\phi}\left(\frac{\rho_f}{M_X}\right)^{1/3}\right]^{1/p} 
\;, \label{ThermalizationTimeNRl}
\end{equation}
where $c_X$ is the fraction of the inflaton energy which initially was
deposited into the field $X$ (this fraction should be measured at the time
when the self-similar evolution starts).  In the present case $p=1/2$, see
Eq. (\ref{p-Mkindep-nonrel}), and similarly to Eq. (\ref{tauTerm}) we find for
the relaxation time
\begin{equation}\label{tauTermNR}
\tau \sim \left(c_X \,\frac{M_{\rm Pl}^{2}}{M_\phi M_X} \right)^{2/3}
\sim \left[ c_X \frac{M_\phi}{ M_X} \right]^{2/3} 10^{8}  \, .
\end{equation}
In this expression $M_\phi/M_X \sim 1$ since Bosons which are much heavier
than the inflaton are not created, and in the opposite regime $X$-particles
would have been relativistic.

As we have seen in Sec.~\ref{sec:NR_expanding} there is no real relaxation of
massive particles when the expansion has become important. If some relaxation
happens, it should occur during the time interval when the scale factor does
not deviate significantly from its initial value. Then
the expansion can be neglected and the relaxation proceeds as in Minkowski
space-time. Let us show that the expressions above agree with the ``naive"
perturbative estimate in the latter case.

\paragraph{Minkowski space-time}

The perturbative relaxation time in the final state can be estimated as
$t_R^{-1} \sim v \sigma n$, where $\sigma \sim \lambda_X^2/M_X^2$ and $n \sim
k_f^3$. Therefore $t_R^{-1} \sim \lambda_X^2 k_f^4 /M_X^3 $. On the other
hand, the rate in the initial state is given by a similar expression, but is
multiplied by large occupation numbers in the initial state
\cite{Tkachev:1991ka} (which can be viewed as Bose-amplification factor),
$\Gamma \sim v \sigma n n_k \sim \lambda_X^2 n^2 /k_i^2 M_X^3$, where we used
$n \sim k_i^3 n_k $. We obtain $t_R \Gamma \sim n^2 / k_f^4 k_i^2 \sim
(\rho/M_X)^{2/3}/k_i^2$, where $\rho = M_X n$. This agrees with
Eq. \eqref{ThermalizationTimeNRl}.

\paragraph{Friedmann universe}

To estimate the thermalization time and temperature we need to know the
typical rate of reactions and the value of the Hubble parameter at the
beginning of self-similar evolution. For definiteness we consider the
situation which arises after preheating in the massive inflaton model coupled
to a heavy field $X$. We assume that self-coupling of the $X$-field is
sufficiently large, such that the ``parametric'' decay of the inflaton is
halted by $X$-rescattering on each other. Using results of
Ref. \cite{Khlebnikov:1997zt} we conclude that at the moment when the inflaton
zero-mode decays completely, the energy density in the $X$-field can be
estimated as $\rho_\chi \sim M_X^4 /\lambda_\chi$, while the rate of
re-scattering is $\eta_0^{-1} \sim M_X$. This gives $H_0 \eta_0 \sim
\lambda_\chi^{-1/2} M_X/M_{\rm Pl} \sim q^{-1/2}$, where $q$ is the initial
resonance parameter. Since $q$ can be very large, the product $H_0 \eta_0$ can
be small and the expansion is not significant at the initial stage of the
self-similar evolution. On the other hand, the time needed to reach the
quantum regime is of order $\tau^{th} \sim (k_f/k_i)^{1/p}$. Since particle
number conserves during the period of self-similar evolution we have $k_f^3
\sim M_X^3/\lambda_\chi$, while $k_i \sim M_X$ at the end of parametric
resonance stage. This gives $\tau^{th} \sim (1/\lambda_\chi)^{1/3p} \sim
\lambda_\chi^{-2/3}$.  The condition $H_0 \eta_0 \tau^{th} < 1$ gives
$\lambda_\chi^{7/6} > M_X/M_{\rm Pl}$ as a necessary condition to reach a
thermal state before the freeze-out of distribution functions.  Using
inflationary normalization, we conclude that non-relativistic particles
created in ``parametric resonance'' have a chance to thermalize between
themselves in an expanding universe if $\lambda_\chi > 10^{-5}$.

\subsection{A faster route to thermalization ?}
\label{Thermalization2}

Considering the regime of free turbulence, we have obtained estimates for
the thermalization time which are in agreement with ``naive'' perturbation theory.
It was important in these estimates that the relativistic free turbulence
propagates with $p = 1/7$. This should be true at sufficiently late times,
when all effects related to zero mode become insignificant. However, free
turbulence driven by 3-particle interactions in the presence of a zero mode
evolves with $p = 1/5$. The evolution of the front of particle distributions is
even faster in the case of driven turbulence, when $p = 2/5$. 
If the quantum domain is already reached during one of these two stages
our estimates for thermalization should be changed.

Here we consider the question whether a subsystem of $\chi$-particles can
reach the quantum region in the regime of a stationary turbulence.

\subsubsection{Driven turbulence}

The quantum domain is reached in the regime of driven turbulence if the power law
of the inertial range will extend up to $n_k \sim 1$. In other words, $n_k =
\left(k/k_T\right)^{-s}$ should be valid up to $k=k_T$. Let us consider the
model were the largest coupling is the self-coupling of the $\chi$-field.  The
normalization of $n_k$ can be fixed if we recall that in the region of the
source, $k \sim k_i$, the $\chi$-particle distribution is given by $n_\chi
\sim 1/\lambda_{\chi}$.  This gives $k_T \sim k_i \lambda_{\chi}^{-1/s}$, or
the time needed to reach the quantum region is given by
\begin{equation}
\tau \sim \lambda_{\chi}^{-1/sp} \; ,
\label{t2quantum}
\end{equation}
where we have used Eq.~\eqref{ThermalizationTime}.

On the other hand, the energy in the subsystem of $\chi$-particles grows in
the regime of driven turbulence as $\rho_\chi (\tau) = \tau\, \rho_\chi(1)$,
and should not exceed the total energy stored in the inflaton zero-mode
oscillations. The initial energy can be estimated as $\rho_\chi(1) \sim k_{\rm
res}^4 n_\chi$, where $k_{\rm res} \sim q^{1/4}\omega_\phi$ and $q$ is the
resonance parameter: $q = \lambda_{\phi\chi}\Phi_0^2/M_\phi^2$ in the
$M_\phi^2 \Phi^2$ inflaton model, or $q = \lambda_{\phi\chi}/\lambda_{\phi}$
in the $\lambda \Phi^4$ inflaton model.  This gives $\rho_\chi(1)/\rho_{\rm
tot} \sim \lambda_{\phi\chi}/\lambda_{\chi}$, and we obtain the bound
\begin{equation}
\tau < \lambda_{\chi}/ \lambda_{\phi\chi} \; .
\label{t2quantum-bound}
\end{equation}
We conclude that the quantum domain can be reached in the regime of driven
turbulence if $\lambda_\chi > \lambda_{\phi\chi}^{sp/(sp+1)} =
\lambda_{\phi\chi}^{3/8} \sim 10^{-4}$.  Here we have used $s=3/2$, $p=2/5$
and $\lambda_{\phi\chi} \sim 10 \lambda_{\phi}$.  These values are realistic,
therefore, physical implications of driven turbulence in applications to
thermalization deserve further study.

\section{Conclusions}
\label{conclusions}

We have studied the process of thermalization of classical systems, which at
some point in their evolution are in a highly non-equilibrium state with
energy being concentrated in a deep ``infra-red'' region of momenta. Such
states naturally appear e.g.  during reheating of the Universe after
cosmological inflation.  We have shown that the process of relaxation in such
systems can be divided, in the general case, into three distinct stages.

In the models of the type we have considered in this paper, the initial stage
of preheating \cite{Kofman:1994rk} is powered by  \textit{parametric
resonance}.  During this initial linear stage the rate of energy transfer is
the fastest. The energy in particles grows exponentially. However, in the
physical situation of reheating after inflation, the coupling constants have
to be sufficiently large to insure an acceptably short time-scale of the
subsequent thermalization, while with large couplings, only a negligible
fraction of the initial inflaton energy is transfered into fluctuations during
the parametric resonance stage \cite{Khlebnikov:1997zt,Prokopec:1997rr}.

We have shown that in such situations the linear stage is followed by the
regime of a \textit{driven stationary turbulence}. During this stage, the
energy in particles grows linearly in time. The regime of stationary
turbulence stops as soon as the energy in particles starts to dominate the
overall energy balance. Therefore, this regime is a major mechanism of energy
transfer from the oscillating inflaton zero-mode into other species in
realistic models of the type we have considered here.  This period of evolution
is also prompt. It should be noted that the source which drives the turbulence
is powerful because coherence effects are still strong in the relevant region
of momenta.

The subsequent long stage of \textit{thermalization} classifies as
\textit{free turbulence}.  This stage should be generic. The energy in
particles is conserved during this epoch, while the shape of the particle
distribution function changes in a self-similar way with the front of the
distribution propagating into the ultra-violet. This stage continues until the
quantum regime is reached and particles can relax to Bose-Einstein
distributions. Applying conventional kinetic theory we have calculated
analytically the time needed to equilibrate a system and the resulting
temperature in terms of coupling constants and initial inflaton amplitude. The
result coincides parametrically with the ``naive'' perturbative estimates
\cite{lindebook} .

We made a comparison of kinetic theory with the numerical integration of
scalar field models on the lattice. We show that, at late times, the kinetic
approach is applicable, resulting in a weak wave turbulence regime
\cite{ZakBook}.  In the models considered numerically, the evolution is driven
by three-particle scattering in the background of  zero-mode oscillations. The
characteristic exponents calculated within the framework of wave kinetic
theory are consistent with the results of our lattice simulations.

\begin{acknowledgments}
We are grateful to S. Khlebnikov, A. Linde, A. Riotto, V. Rubakov, C. Schmid,
D. Semikoz and M. Shaposhnikov for many useful discussions during various
stages of this project. R.M. thanks the Tomalla Foundation for financial
support.
\end{acknowledgments}

\appendix

\section{Numerical Procedure}
\label{NumericalProcedure}

In our study we have developed and employed a higher accuracy version of the
LATTICEEASY code \cite{Felder:2000hq}. Various correlation functions were
measured with the use of Fast Fourier Transform (FFT), adopted from {\it
Numerical Recepies} \cite{NumRec}. Essential details of our procedure can be
found in Refs. \cite{Khlebnikov:1996mc,Khlebnikov:1997zt,Khlebnikov:1997di,%
Felder:2000hq}.  Here we describe specific choices of parameters, significant
important differences in the integration scheme and give exact definitions of
lattice observables.

The numerical integration was done on a 3-D cubic lattice with periodic
boundary conditions.  The lattice is parameterized by the box size $L$, and
the number of lattice points per dimension, $N$.  These give the lattice
spacing $b \equiv L/N$ and the total number of lattice sites, $N^3$ in three
dimensions.

The results presented in the paper are taken from simulations with $256^3$
lattice sites and a box size $L$ chosen to fit a particular problem. For
example, in the case of Eq. \eqref{BEq}, $L=7.5\pi$.  With this box size the
infrared modes which belong to the resonance band are still well represented,
while the ultraviolet lattice cut-off is sufficiently far away from the
occupied modes, therefore the particle spectra are not distorted even at late
times.  We have studied the dependence of our results on the lattice- and the
box size to avoid lattice artifacts.

The finite-differences scheme that was used is 2nd order in time and 4-th
order in space.

\subsubsection{Finite-differences scheme}

We write the equations of motion \eqref{BEq} or \eqref{eq_mot_M:phi},
\eqref{eq_mot_M:chi} as fourth order finite differences on a three-dimensional
spatial cubic lattice with periodic boundary conditions. The corresponding
equations were evolved with the use of a symplectic integration scheme. Details
are as follows.

Particle wave numbers are discrete on the lattice,
$\bm{k}=(n_{1},n_{2},n_{3})k_{\mathrm{0}}$, where $-N/2\leq n_{j}\leq N/2$ and
$k_{0}=(2\pi)/L$. The phase space is restricted to $k_0 \leq k \leq k_{\rm
max}$, where $k_{\rm max} = \sqrt{3}k_{0}N/2$. To avoid distortion at high
momenta, it is desirable to take large N. This, however, is limited by the
capabilities of the computer used. The choice of small values for $L$ is
also prohibited since that will lead to infrared distortions and may even move
the resonance band out of the integration box. The problem is alleviated by
the choice of a finite-differences scheme which is fourth order in space. This
can be quantified in the following way.

\begin{figure}[t]
\includegraphics*[width=3.9in]{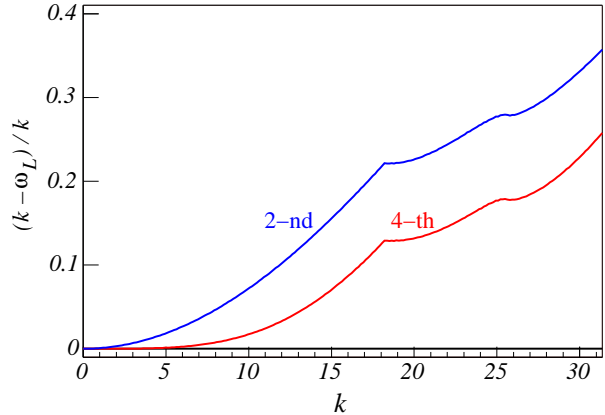}
\caption{Deviation of the dispersion law for massless excitations on the
  lattice, $(k - \omega_L)/k$,  for  second and fourth order
  finite-difference schemes.
\label{LatOmega}}
\end{figure}

The lattice realization of the Laplacian in our scheme is given by:
\begin{widetext}
\begin {equation}\label{def:LatLap}
\bigtriangleup_{\mathrm{L}} \Phi(\bm{x}) = \frac{1}{b^{2}}
\sum_{\bm{e}} \left[-\frac{1}{12}\Phi(x+2b\bm{e})+\frac{4}{3}
\Phi({x+b\bm{e}})
-\frac{5}{2}\Phi(x)+\frac{4}{3}\Phi(x-b\bm{e})-\frac{1}{12}
\Phi({x-2b\bm{e}})
\right]\, .
\end{equation}
\end{widetext}
The vector index $\bm{e}$ runs over the three orthonormal directions of the
lattice.  $\bigtriangleup_{\mathrm{L}}$ is a fourth order approximation, i. e.
$(\bigtriangleup-\bigtriangleup_{\mathrm{L}})\Phi(\bm{x})\sim
\mathcal{O}(b^4)$ for a differentiable function $\Phi(\bm{x})$ .  The Fourier
transform of $ \bigtriangleup_{\mathrm{L}} $ differs from that of $
\bigtriangleup $, which would be given by multiplication with $k^{2}$.
Therefore the dispersion relation for a massless field on the lattice is also
different, and is given by
\begin{equation}\label{kL}
 \omega^{2}_{\mathrm{L}}(k) =  \frac{1}{b^{2}} \sum_{i=1}^{3} 
\left( \frac{5}{2} - \frac{8}{3} 
        \cos(b k_i) + \frac{1}{6} \cos(2 a k_i) \right) ,
\end{equation} 
We find that $\omega^{2}_{\mathrm{L}} \leq k^2$ and $\omega_{L}^2-k^2\sim
\mathcal{O}(k^6b^4)$ for small $k$.  Numerically, for $k\leq k_{\rm max}/3$,
the relative difference between $k$ and $\omega_L$ is less then a percent,
while for larger $k$ it grows up to about 30 percent difference at $k=k_{\rm
max}$.  This means, that we can expect essentially undistorted self-similar
and turbulent solutions on the lattice, if the dominating modes have wave
vectors $k\lesssim k_{\rm max}/3$.  In the case of a second order realization
of $\bigtriangleup_{\mathrm{L}}$, we find a considerably smaller available
phase space, $k\lesssim k_{\rm max}/10$. This is illustrated in
Fig.~\ref{LatOmega} where we plot $(k - \omega_L)/k$ as a function of $k$ for
the second and fourth order calculation schemes on the lattice $L=7.5\pi$,
used in our integration of the problem Eq. \eqref{BEq}. We see that up to $k =
10$, which essentially encompasses the support region of the distribution
functions, see Fig.~\ref{self_sim}, the dispersion law on the lattice
represents the continuum correctly. That is why self-similarity was not
distorted on our lattice and could have being detected. (The small deviations
from self-similarity, which can be observed at the very tail of the
distribution and at the latest time, see Fig.~\ref{self_sim}, are caused by the
distortion of the dispersion law which starts to be non-negligible here).

\subsubsection{Classical Approximation and Stochastic Initial Conditions}
The initial linear stage of parametric resonance has a complete quantum
description, which is best expressed in the language of Bogoljubov
transformations.  The quantum description of this linear problem can be mapped
into an equivalent classical problem \cite{Khlebnikov:1996mc}.  In our
dimensionless variables the initial conditions for the classical description
are given by the following probability distribution for field fluctuations in
Fourier space:
\begin{equation}\label{ContStochInCon}
\begin{split}
\mathbf{P}[\psi,\dot\psi]\;\sim\;&\exp\left\lbrace
-\frac{2}{\lambda_{\Phi}}\int  d^3k\,  2\omega_{k}^{\psi}(\tau_0)\,
|\psi_{\bm{k}}|^2 \right\rbrace\,\\&\times\delta^{F}
\left( \dot\psi_{\bm{k}}+i\omega_{k}^{\psi}(\tau_0)\psi_{\bm{k}}\right) 
\end{split}
\end{equation} 
Here "$\psi $" should be replaced by one of the fields $\phi $ or $ \chi $
that are the dynamical variables in the simlated equations \eqref{BEq} or
\eqref{eq_mot_M:phi}, \eqref{eq_mot_M:chi}, and
$\omega_{k}^{\phi}(\eta_0)\simeq\sqrt{k_{L}^2+3}$, while
$\omega_{k}^{\chi}(\eta_0)\simeq\sqrt{k_{L}^2+g}$.  The symbol
$\delta^F\left(\ldots \right)$ is a functional Dirac-distribution so that the
canonical momenta are locked to the classical trajectory.

\subsubsection{Measured Quantities:}

We measure various physical quantities both in configuration space and in
Fourier space. In configuration space it is convenient to measure the zero
mode, $\phi_0 \equiv \langle\phi\rangle$, and the variance, $var(\phi) \equiv
\langle\phi^2\rangle -\phi_0^2$. In Fourier space we measure particle number
and other correlators.

For large $N$ lattice averages basically coincide with the statistical
ones (ergodic theorem). We use this fact to measure expectation values
(zero-modes), variances and higher cumulants of fields and their
conjugate momenta.

\paragraph*{Spatial Lattice Averages:}

For averages defined in configuration space $\langle {\cal O} \rangle \equiv
V^{-1}\int d^{3}x\, {\cal O}$, which on the lattice is expressed as the sum
over the lattice points $\langle {\cal O} \rangle \equiv N^{-3}\sum_i {\cal
O}_i$.

\paragraph*{Fourier Spectra:}

For monitoring purposes we make a FFT transform at least every period of
inflaton oscillation. The wave amplitudes of fourier transformed fields are
defined by Eq.~\eqref{micro_a_eq}, see Appendix \ref{WavesKinetics}.  In the
dimensionless units that we use in the numerical simulation the physical wave
amplitudes take the form
\begin{equation}\label{def:waveampl}
\tilde{a}^{\psi}_{\bm k} \equiv  \frac{1}{\sqrt{\lambda_{\Phi}}}\;
\frac{\omega^{\psi}_k \psi_{\bm k} +
i \dot{\psi}_{\bm k}}
{{(2\pi)^{3/2}}\sqrt{2\omega^{\psi}_k}} \;, 
\end{equation}
where again "$\psi $" stands for the dynamical variables $ \phi $ or $\chi $
in the equations \eqref{BEq} or \eqref{eq_mot_M:phi}, \eqref{eq_mot_M:chi}.
The dimensionless frequencies are given by $\omega^{\psi}_k \equiv
\sqrt{k_{L}^2+{m^{\psi}_{\mathrm{eff}}}^2}$, where
${m^{\phi}_{\mathrm{eff}}}^2=3\langle\varphi^2\rangle+g\langle\chi^2\rangle$
and
${m^{\chi}_{\mathrm{eff}}}^2=g\langle\varphi^2\rangle+3h\langle\chi^2\rangle$.
Making use of $a_k$, we calculate various correlators, $n(k) \equiv \langle
a^*_{\bm k} a_{\bm k}\rangle$, $\sigma (k) \equiv \langle a_{\bm k} a_{-\bm
k}\rangle$, $\langle a^* a^* a a\rangle$, etc. The first one, which
corresponds to the particle occupation numbers, is of prime interest.

Note that with this simple definition of quasi-particles the Hamiltonian is
not diagonal in terms of $a^\chi_{\bm k}$ and $a^\phi_{\bm k}$ wave amplitudes
if interaction energy is important.  Therefore, the related definition of e.g.
particle number is good only for modes with dominating kinetic energy.

\section{Kinetic equation for classical waves}
\label{WavesKinetics}

Following the general approach of Refs.~\cite{ZakBook,zak:85} we derive the
wave kinetic equation for the classical system of interest, the massive $
\lambda\phi^4 $- theory in $ d $ dimensions with Hamiltonian density
\begin{equation}\label{scalar-field-ham}
\mathcal{H}=\frac{1}{2}\dot\phi^{2}+\frac{1}{2}(\nabla\phi)^{2} +
\frac{M^{2}}{2}\phi^{2}+\frac{\lambda}{4} \phi^{4} \; ,
\end{equation} 
and in the presence of an oscillating classical background.  
We assume random
wave fields  which are statistically uniform, i.e.  the equal time correlation
functions of $ \phi $ and its canonical conjugate momentum $ \dot\phi $ are
homogeneous and isotropic. We also assume the field to be weakly interacting.

The first step in the derivation of the kinetic equation for an arbitrary
system is to find Fourier wave amplitudes, $a_{\bm{k}}$, such that the
quadratic part of the Hamiltonian is diagonal in $a_{\bm{k}}$, i.e. :
\begin{equation}
{H}_2=\int d^d k \, \omega_k a^*_{\bm k} a^{}_{\bm k} \; .
\end{equation} 
The general equation of motion for the wave amplitudes is
\begin{equation} \label{eq-mot-wampl}
\frac{d a(k,t)}{d t} = \frac{\partial a(k,t)}{\partial t}
-i \omega_k a(k,t)  - i 
\frac{\delta H_{int}}{\delta a^*(k,t)}, 
\end{equation} 
where $H_{int} \equiv H - {H}_2$.  The first term in the l.h.s. is due to a
possible explicit time-dependence in the definition of $a_{\bm{k}} $, which
can appear for example in case of a time-varying background.

In the kinetic approach we want to get rid of rapidly varying phases of the
wave amplitudes, i.e. to derive the equation for the slowly changing
``occupation numbers'', $n^{~}_k \sim a^*_ {\bm k}a^{~}_{\bm k}$. To achieve
this we multiply Eq.~\eqref{eq-mot-wampl} by $a^*_{\bm k}$, subtract the
complex-conjugate expression and average.  The result will contain higher
order correlators induced by interaction terms. The resulting BBGKY-hierarchy
of equations for Fourier cumulants can be solved, e.g., in the random phase
approximation in consistent perturbative expansion.

In the case of \eqref{scalar-field-ham} the wave amplitudes for the fluctuation
fields are solutions of
\begin{eqnarray}\label{def:ak:inverted}
\delta\phi^{~}_{\bm{k}}&=&\frac{(2\pi)^{d/2}}{\sqrt{2\omega_{k}}}
\left( a^{~}_{\bm{k}}+a_{-\bm{k}}^{*}\right)\\
\delta\dot\phi^{~}_{\bm{k}}&=&\frac{(2\pi)^{d/2}\sqrt{\omega_{k}}}{\sqrt{2} i}
\left( a^{~}_{\bm{k}}-a_{-\bm{k}}^{*}\right)
\end{eqnarray}
where $\delta\phi_{\bm k}$ and $\delta\dot{\phi}_{\bm k}$ are Fourier
transforms of the canonical field and of its conjugate momenta respectively,
shifted by the ``zero mode'' $\phi_{0}=\langle\phi\rangle$ and
$\dot\phi_{0}=\langle\dot\phi\rangle$. This gives
\begin{equation}\label{def:ak}
{a}^{~}_{\bm k}\equiv  
\frac{\omega^{~}_k \delta\phi^{~}_{\bm k} +
i \delta\dot{\phi}^{~}_{\bm k}}
{{(2\pi)^{d/2}}\sqrt{2\omega^{~}_k}} \; .
\end{equation}
From the start we include in $\omega_{k}$ the interaction with the bath of
fluctuations,
\begin{equation}\label{ok:def} 
\omega_{k}^{2}= k^2+M^2+3\lambda\phi_{0}^{2}+ 
3\lambda\langle\delta\phi^{2}\rangle ,\;
\end{equation}
i.e. $a_{\bm k}$ correspond to ``quasiparticles''. The second order
correlators in homogeneous and isotropic background should be ``diagonal''

\begin{eqnarray}
\langle a_{\bm{k}}^{*} a_{\bm{q}^{~}}\rangle &=& n^{~}_{k}
\delta^{(d)}(\bm{k}-\bm{q})\\
\langle a_{\bm{k}} a_{\bm{q}}\rangle 
&=& \sigma^{~}_{k}\delta^{(d)}(\bm{p}+\bm{q})
\end{eqnarray} 

\subsection{Microscopical Equations of Motion}

We derive equations of motion for the zero mode and for wave amplitudes
starting from
\begin{equation}\label{eqnmot:}  
\square\phi+M^2\phi+\lambda\phi^{3}=0 \; .
\end{equation} 

\paragraph{Zero-Mode.}

Averaging \eqref{eqnmot:} we obtain 
  \begin{equation}\label{micro_zm_eq} 
\ddot\phi_{0}+
(M^2+3\lambda\langle\delta\phi^{2}\rangle)\phi_{0} +
\lambda\phi_{0}^3+\lambda\langle\delta\phi^3\rangle = 0 \; .
\end{equation} 
In this equation $\langle\delta\phi^3\rangle$ is small compared to the other
terms and may be neglected locally in time in a state which is close to a
Gaussian.  If additionally $\langle \delta\phi^2 \rangle$ is either weakly
varying or small compared to all other terms in \eqref{micro_zm_eq} the
solution is given by the Jacobian cosine $cn $ \cite{Greene:1997fu,Abramowitz,
Bateman:53}
\begin{equation}\label{zm:Jacobi-app} 
\phi_{0}(t)\simeq  \bar{\phi}_{0}\;\mathrm{cn}\left(\mu t, 
\frac{1}{\sqrt{2}}\frac{\lambda \bar{\phi}_{0}}{\mu}\right) \; ,
\end{equation} 
where $ \bar{\phi}_{0} $ is the amplitude and
\begin{equation}
\mu \equiv \sqrt{\lambda \bar{\phi}_{0}^{2}+
3\lambda\langle\delta\phi^{2}\rangle+M^2} \; .
\end{equation} 
The period of this function is 
\begin{equation}
T_{0}=4\mu^{-1}\,K\left( \frac{1}{\sqrt{2}}
\frac{\lambda^{1/2} \bar{\phi}_{0}}{\mu}\right) \; ,
\end{equation} 
where $K(y)$ is the complete elliptic integral of the first kind.
This defines the effective frequency, $\omega_c = 2\pi /T_{0}$. 
In the large amplitude limit, $\mu = \lambda^{1/2} \bar{\phi}_{0}$, we
find $ \omega_{c}\simeq 0.85\mu $. For arbitrary $\mu$ one can write the
following decomposition 
\begin{equation}\label{omega_c:approx} 
\omega_c\simeq \mu-\frac{1}{8}\frac{\lambda \bar\phi_0^{2}}{\mu} \; ,
\end{equation}
which is fairly accurate, the maximum deviation from the exact expression is
3\% at $\mu = \lambda^{1/2} \bar{\phi}_{0}$. For small amplitude of zero-mode
oscillations, this expression can be further approximated as
\begin{equation}\label{omega_c:approx2} 
\omega_{c}\simeq M_{\rm eff}\left( 1+\frac{3\lambda\bar\phi_{0}^2}
{8M_{\rm eff}^{2}}\right), 
\end{equation}
where $M_{\rm eff} \equiv (M^2+3\lambda\langle\delta\phi^{2}\rangle )^{1/2} $.
This deviates from exact result by less than 4\% at $\lambda \bar\phi_{0} <
M_{\rm eff}$.

For a general discussion of the kinetic equations in
the background of a zero mode it might be useful to expand ${\phi}_{0}(t) $
in a Fourier series. However, this decomposition for the elliptic Jacobi
function is strongly dominated by the first harmonic with $\omega = \omega_c$.
Even at $\mu = \lambda^{1/2} \bar{\phi}_{0}$, the relative amplitude
of the first harmonic is $\approx 0.96$, and it approaches unity with
decreasing $\bar{\phi}_{0}$. Therefore, in what follows we will restrict
ourselves to the first term in the Fourier decomposition of ${\phi}_{0}(t)$.

It is useful to define wave amplitudes for the zero-mode
\begin{equation}
a_{c}\equiv \sqrt{2\omega_{c}}\, \bar\phi_{0}e^{-i\omega_{c}t} 
\end{equation} 
in terms of which the zero-mode can be represented as
\begin{eqnarray}
\phi_{0}=\frac{a_{c}+a_{c}^{*}}{\sqrt{2\omega_{c}}} \; .
\end{eqnarray} 
One can also introduce an effective occupation number of "condensed waves",
\begin{equation}
n^{}_{c}=a_{c}^{*}a^{}_{c}=2\omega_{c}\bar\phi_{0}^{2}\; .
\end{equation}

\paragraph{Wave amplitudes.}

The equations of motion for the wave amplitudes with non-zero momentum can be
written as
\begin{equation}\label{micro_a_eq} 
\begin{split}
\dot{a}_{\bm{k}}=-i\omega_{k}{a}_{\bm{k}}+
\frac{1}{2}\frac{\dot\omega_{k}}{\omega_{k}}{a}_{-\bm{k}}^{*}
+C^{(3)}_{\bm{k}} + C^{(4)}_{\bm{k}} \:, 
\end{split}
\end{equation}
where
\[
\begin{split}
C^{(3)}_{\bm{k}} \equiv &-3i\lambda\phi_{0}\int d\Omega_{k12}
\left[\delta\phi_{\bm{p}_{1}} \delta\phi_{\bm{p}_{2}}\right.\\
&-\left.
\langle\delta\phi_{\bm{p}_{1}}
\delta\phi_{\bm{p}_{2}}\rangle \right] 
\delta^{(d)}(\bm{k}-\bm{p}_{1}-\bm{p}_{2}) \\
C^{(4)}_{\bm{k}} \equiv &-i{\lambda} \int d\Omega_{k123}
\left[\delta\phi_{\bm{p}_{1}}
\delta\phi_{\bm{p}_{2}}\delta\phi_{\bm{p}_{3}}
-3\delta\phi_{\bm{p}_{1}}\langle
\delta\phi_{\bm{p}_{2}}\delta\phi_{\bm{p}_{3}}\rangle \right.\\
&-\left.
\langle\delta\phi_{\bm{p}_{1}}
\delta\phi_{\bm{p}_{2}}\delta\phi_{\bm{p}_{3}}\rangle \right]
\delta^{(d)}(\bm{k}-\bm{p}_{1}-\bm{p}_{2}-\bm{p}_{3})\;,
\end{split}
\]
and
\[
d\Omega_{k12}\equiv
\frac{d^dp_1d^dp_2}{\sqrt{2\omega_{k}}(2\pi)^{3d/2}}\;,
~~~~~
d\Omega_{k123}\equiv
\frac{d^dp_1d^dp_2d^dp_3}{\sqrt{2\omega_{k}}(2\pi)^{5d/2}}
\]
$ C^{(3)}_{\bm{k}} $ describes three wave interactions in the background of a
zero-mode-and, while $C^{(4)}_{\bm{k}} $ corresponds to four wave scattering.
The averages in these expressions appeared because, first, we separated the
zero-mode out of the equation for fluctuations, and, second, we used the
effective frequency for quasiparticles, Eq. \eqref{ok:def}.  Due to this
choice the averages of $C^{(a)}$ times $a_{\bm k}$ or $a^{*}_{\bm k}$ will
have the structure of cumulants, which in turn will deviate from zero only due
to correlations induced by processes of scattering.
 
Multiplying \eqref{micro_a_eq} by $a^*_{\bm k}$ or by $a_{\bm k}$ and
subtracting the complex-conjugate expressions, we find
\begin{eqnarray}\label{eq-for-n} 
\dot n_{k}&=&\frac{\dot\omega_{k}}{\omega_{k}}\mathrm{Re} \, \sigma_{k}+
\mathrm{Im} I_{3}(k)+ \mathrm{Im} I_{4}(k) \\ 
i\dot\sigma_{k}&=&2\omega_{k}\sigma_{k} +
\frac{i}{2}\frac{\dot\omega_{k}}{\omega_{k}}n_{k}
+I_{3}^*(k)+I_{4}^*(k)
\label{eq-for-sigma} 
\end{eqnarray} 
where
\begin{eqnarray}
I_{3}{(k)}&=&6{\lambda\phi_{0}}\int d\Omega_{k12}\,
\langle a_{\bm{k}}^{*} \delta\phi_{\bm{p}_{1}}
\delta\phi_{\bm{p}_{2}}\rangle_{c} \label{I3:micr} \\
 I_{4}{(k)}&=&2{\lambda}\int d\Omega_{k123}\,
\langle a_{\bm{k}}^{*} \delta\phi_{\bm{p}_{1}}
\delta\phi_{\bm{p}_{2}}\delta\phi_{\bm{p}_{3}}\rangle_{c}\qquad 
\label{I4:micr} 	
\end{eqnarray}
Here $\langle\dots\rangle_{c} $ denotes the cumulants, which in a diagrammatic
language are identified with \textit{connected} diagrams, see
e.g. Ref.~\cite{ParisiBook}.

In situations when $I_{3}(k)$ and $I_{4}(k)$ are negligible,
Eqs. \eqref{eq-for-n} and \eqref{eq-for-sigma} describe particle creation in a
time-dependent classical background, or parametric excitation if it is
periodic.  (Note that the quantum version of these equations at this stage can
be obtained by the formal substitution $n_k \rightarrow 1/2+n_k$.)

Note the following
\begin{itemize}
\item In our case $\omega_{k}$ contains an rapidly oscillating term, due to
  interaction with the zero mode. However, at late times and at large $k$ it
  is small. e.g., in our numerical integration in the region of $k$ near the
  peak of the spectral energy distribution, this term is of order $10^{-3}$,
  see Figs~\ref{spat_av} and \ref{self_sim}. We neglect this term in what
  follows.
\item The coefficient in front of the integral Eq. \eqref{I3:micr} is rapidly
  oscillating. Moreover, oscillations are not harmonic if the amplitude of
  $\phi_0$ is large. Unharmonicity can be dealt with by expanding $\phi_0(t)$
  in Fourier time series and considering each of the terms separately.  We
  restrict ourselves to the leading harmonic in this expansion since at late
  times the unharmonicity is small.
\item The cumulants contain different combinations of $a_{\bm k}$ and
  $a^*_{\bm k}$, see Eqs.~\eqref{def:ak:inverted}. It is well known that the
  leading contribution to the resulting kinetic equation is due to the
  "resonant wave interactions", or, in the language of particle physics, only
  those terms survive, which are on the ``mass-shell''.  In our case those
  will be $\langle a_{\bm{p}}^{*}
  a_{\bm{p}_{1}}^{}a_{\bm{p}_{2}}^{}\rangle_{c} $ and $\langle a_{\bm{p}}^{*}
  a_{\bm{p}_{1}}^{*} a_{ \bm{p}_{2}}^{}\rangle_{c} $ for interactions which
  involve the zero mode, and $\langle a_{\bm{p}}^{*}
  a_{\bm{p}_{1}}^{*}a_{\bm{p}_{2}}^{}a_{\bm{p}_{3}}^{}\rangle_{c} $ otherwise.
  We restrict our attention to these cumulants only.
\item We neglect ``anomalous'' correlators, $\sigma_k$. These are small in the
  inertial range of turbulence as our lattice calculations show, but may be
  important otherwise.
\end{itemize}

\subsection{Leading Asymptotic of Collision Terms in Kinetic Approximation}

For a free random field the cumulants Eqs. \eqref{I3:micr} and \eqref{I4:micr}
are zero, and $\dot{n}_k=0$ to the first order in perturbation theory.  To
calculate $\dot{n}_k$ in second order one has to know the solutions for
cumulants in the first order with respect to interactions.
 
We use the equation of motion for wave amplitudes, Eq. \eqref {micro_a_eq}, to
calculate the time derivatives of the cumulants, $\partial_{t}
a_{c}^{*}\langle a_{\bm{p}}^{*} a_{\bm{p}_{1}}^{}a_{\bm{p}_{2}}^{}\rangle_{c}
$ and $\partial_{t}\langle a_{\bm{p}}^{*}
a_{\bm{p}_{1}}^{*}a_{\bm{p}_{2}}^{}a_{\bm{p}_{3}}^{}\rangle_{c} $.  Higher
order correlators which appear in this procedure can be used in zeroth order
of perturbation theory, i.e. they can be decomposed in $n_k$ assuming
Gaussianity.  To simplify the equations we use the following notations for
products of $n_k$ which appear in these decompositions
\begin{eqnarray}
\mathrm{F}^{\,\bm{p}_{\,\, }}_{\,\bm{p}_{1}\bm{p}_{2}} &\equiv&
n_{c} n_{\bm{p}_1} n_{\bm{p}_2}- n_c n_{\bm{p}}
\left( n_{\bm{p}_{1}}+ n_{\bm{p}_{2} }\right) 
\label{ClassicalF3} \\
\mathrm{F}^{\,\bm{p}_{\,\, }\bm{p}_{1}}_{\,\bm{p}_{2}\bm{p}_{3}}&\equiv& 
\left( n_{\bm{p}}+n_{\bm{p}_{1}} \right)n_{\bm{p}_{2}}n_{\bm{p}_{3}} -
n_{\bm{p}} n_{\bm{p}_{1}}\left( n_{\bm{p}_{2}}+n_{\bm{p}_{3}}\right)\qquad 
\label{ClassicalF4} 
\end{eqnarray}
We find, keeping the terms which will have resonant behavior
\begin{widetext}
\begin{eqnarray}
\partial_{t}\, a_{c}^{*}\langle a_{\bm{p}}^{*} 
a_{\bm{p}_{1}}^{}a_{\bm{p}_{2}}^{}\rangle_{c} & \;\simeq\;&i(\omega_{c}+
\omega_{p}-\omega_{p_{1}}-\omega_{p_{2}})
 \,a_{c}^{*}\langle a_{\bm{p}}^{*} 
a_{\bm{p}_{1}}^{}a_{\bm{p}_{2}}^{}\rangle_{c} \;+\;
\frac{i 6 \lambda \; \delta^{(d)}(\bm{p}- \bm{p}_{1}- \bm{p}_{2})
}{{(2\pi)^{d/2}
\sqrt{2\omega_{c}2\omega_{p}2\omega_{p_{1}}2\omega_{p_{2}}}}}\;
\mathrm{F}^{\,\bm{p}_{\,\, }}_{\,\bm{p}_{1}\bm{p}_{2}}\label{firstterm_dot} \\
 \partial_{t}\,\langle a_{\bm{p}}^{*} a_{\bm{p}_{1}}^{*}
a_{\bm{p}_{2}}^{}a_{\bm{p}_{3}}^{}\rangle_{c}& \simeq &
i(\omega_{p}+\omega_{p_{1}} - \omega_{p_{2}} -\omega_{p_{3}})\,
\langle a_{\bm{p}}^{*} a_{\bm{p}_{1}}^{*}a_{\bm{p}_{2}}^{}
a_{\bm{p}_{3}}^{}\rangle_{c} \;+\; \frac{i 6 \lambda\; \delta^{(d)}
(\bm{p}+ \bm{p}_{1}- \bm{p}_{2}- \bm{p}_{3})}{(2\pi)^{3d/2}
\sqrt{2\omega_{p}2\omega_{p_{1}}2\omega_{p_{2}}2\omega_{p_{3}}}}\;
\mathrm{F}^{\,\bm{p}_{\,\, }\bm{p}_{1}}_{\,\bm{p}_{2}\bm{p}_{3}}
\label{secondterm_dot}
\end{eqnarray} 
\end{widetext}
These equations have the common structure, $i\dot{J} = \Delta \omega J -
A$. Since $A$ corresponds to the zeroth order in perturbations, it assumed to
be time independent here. An Appropriate particular solution for cumulants is
therefore given by $J = A/(\Delta \omega + i\epsilon)$, see e.g.
Ref.~\cite{ZakBook}. Using the relation $\mathrm{Im}(x + i\epsilon)^{-1} =
-\pi \delta(x)$ we obtain
\begin{widetext}
\begin{eqnarray}
\mathrm{Im}\,a_{c}^{*}\langle a_{\bm{p}}^{*} a_{\bm{p}_{1}}^{~}
a_{\bm{p}_{2}}^{~}\rangle _{c}
&\;\simeq\;&   
3\lambda\;\frac{\delta^{(d)}( \bm{p}-\bm{p}_{1}- \bm{p}_{2})
\delta(\omega_{c}+\omega_{p}-\omega_{p_{1}}-\omega_{p_{2}})}
{{(2\pi)^{d/2-1}\sqrt{2\omega_{c}2\omega_{p}2\omega_{p_{1}}2
\omega_{p_{2}}}}}\;\mathrm{F}^{\,\bm{p}_{\,\, }}_{\,\bm{p}_{1}\bm{p}_{2}}
\label{firstcumul}\\ 
\mathrm{Im}\,\langle a_{\bm{p}}^{*} a_{\bm{p}_{1}}^{*}
a_{\bm{p}_{2}}^{~}a_{\bm{p}_{3}}^{~}\rangle_{c}&\; \simeq\; &
 3\lambda\;\frac{\delta^{(d)}( \bm{p}+ \bm{p}_{1}- \bm{p}_{2}- \bm{p}_{3})
\delta(\omega_{p} +\omega_{p_{1}}-\omega_{p_{2}}-
\omega_{p_{3}})}
{(2\pi)^{3d/2-1}\sqrt{2\omega_{p}2\omega_{p_{1}}2
\omega_{p_{2}}2\omega_{p_{3}}}}\;
\mathrm{F}^{\,\bm{p}_{\,\, }\bm{p}_{1}}_{\,\bm{p}_{2}\bm{p}_{3}}
\label{secondcumul} 
\end{eqnarray}
\end{widetext}

\subsection{Isotropic Wave Kinetic Equations}

Applying this result to Eqs.~\eqref{I3:micr} and \eqref{I4:micr} we obtain the
kinetic equation for wave occupation numbers $ n_{k}$
\begin{equation}\label{set_of_kin_eq-part}
\dot n_{k}\;=\; I_{k}^{(3)}\,+\,I_{k}^{(4)} \; ,
\end{equation} 
where
\begin{eqnarray}
I_{k}^{(3)}&=&\int d\Omega^{\,\bm{k}_{\,\, }}_{\,\bm{p}_{1}
\bm{p}_{2}}\mathrm{F}^{\,\bm{k}_{\,\, }}_{\,\bm{p}_{1}\bm{p}_{2}}-2
\int d\Omega_{\,\bm{p}_{2 }}^{\,\bm{k}_{\,\,}\bm{p}_{1}}
\mathrm{F}^{\,\bm{p}_{2 }}_{\,\bm{k}_{\,\,}\bm{p}_{1}}\label{I3:makr} \\
I_{k}^{(4)}&=&\,
\int d\Omega^{\bm{k}_{\,\, }\bm{p}_{1}}_{\,\bm{p}_{2}\bm{p}_{3}}\;
\mathrm{F}^{\,\bm{k}_{\,\, }\bm{p}_{1}}_{\,\bm{p}_{2}\bm{p}_{3}}\label{I4:makr}
\end{eqnarray} 
and
\begin{equation}
\begin{split}
d\Omega^{\bm{k}_{\,\, }}_{\,\bm{p}_{1}\bm{p}_{2}}\;\equiv\;&
\frac{18\lambda^{2}\,d^dp_1d^dp_2\,\delta^{(d)}( \bm{k}- \bm{p}_{1}- 
\bm{p}_{2})}
{(2\pi)^{d-1}2\omega_{c}{2\omega_{k}}2\omega_{p_{1}}2\omega_{p_{2}}}
\\
& \times\delta(\omega_{c}+\omega_{k}-\omega_{p_{1}}-\omega_{p_{2}})
\end{split}
\end{equation} 
\begin{equation}
\begin{split}
d\Omega^{\,\bm{k}_{\,\, }\bm{p}_{1}}_{\,\bm{p}_{2}}\;:=\;&
\frac{18\lambda^{2}\,d^dp_1d^dp_2\,\delta^{(d)}( \bm{k}+ \bm{p}_{1}- 
\bm{p}_{2})}
{(2\pi)^{d-1}{2\omega_{k}}2\omega_{p_{1}}2\omega_{p_{2}}2\omega_{c}}
\\
& \times\delta(\omega_{k}+\omega_{p_{1}}-\omega_{p_{2}}-\omega_{c})
\end{split}
\end{equation} 
\begin{equation}
\begin{split}
d\Omega^{\bm{k}_{\,\, }\bm{p}_{1}}_{\,\bm{p}_{2}\bm{p}_{3}}\;\equiv\;&
\frac{18\lambda^{2}\,d^dp_1d^dp_2 d^dp_3\,\delta^{(d)}( \bm{k}+ \bm{p}_{1}- 
\bm{p}_{2}- \bm{p}_{3})}
{(2\pi)^{2d-1}2\omega_{k}{2\omega_{p_{1}}}2\omega_{p_{2}}2\omega_{p_{3}}}
\\
& \times\delta(\omega_{k}+\omega_{p_{1}}-\omega_{p_{2}}-\omega_{p_{3}})
\end{split}
\end{equation} 

Both terms in \eqref{set_of_kin_eq-part} describe scattering processes of two
waves into two other ones. In \eqref{I3:makr} one of them comes from the
zero-mode, while in \eqref{I4:makr} all four are from the fluctuation field.
Energy conservation in the interactions with the zero-mode, $\omega_{c} +
\omega_{k} - \omega_{p_{1}} - \omega_{p_{2}} = \varepsilon_k -
\varepsilon_{p_{1}}- \varepsilon_{p_{2}} $, can be written as conservation of
the energies
\begin{equation}
\varepsilon^{~}_p \;\equiv\; \omega^{~}_p - \omega^{~}_{c} \;,
\label{rel-kin-en-def}
\end{equation} 
 which
for small zero-mode amplitude (where $\omega_{c}\simeq \omega_{p=0} $)
equals to the  {\it relativistic}
kinetic energy.
Therefore, transport of energy over momentum space should be considered as
transport of kinetic energy in this case.

These relations, Eqs. \eqref{I3:makr} and \eqref{I4:makr},  for 3- and
4-particle collision integrals can be reduced to one and two dimensional
integrations respectively, if the distribution functions are isotropic, for
details see e.g.  Refs \cite{Semikoz:1997rd,ZakBook}.  

The collision integral for 3-particle interactions in $d=3$ takes the form
\begin{eqnarray}
I_k^{(3)} &=& \frac{9\lambda^2 \bar{\phi}_0^2}{16\pi \,\omega_k \,  k} 
\left( \int^{\varepsilon_k}_0 d\varepsilon^{~}_2\,
[n^{~}_3 n^{~}_2 - n^{~}_k\, (n^{~}_3 + n^{~}_2)]\,  \right.\nonumber \\
 &+& \left. 2 \int_{\varepsilon_k}^\infty d\varepsilon^{~}_2 \,
[ n^{~}_2(n^{~}_k + n^{~}_1)-n^{~}_k\, n^{~}_1 ]\,  \right) ~,
\label{I3.0}
\end{eqnarray}
where $\varepsilon_i \equiv \varepsilon(p_i)$ and $n_i \equiv
n(\varepsilon_i)$. Energy conservation in this case corresponds to
$\varepsilon_3 = \varepsilon_k - \varepsilon_2$ and
$\varepsilon_1=\varepsilon_2 - \varepsilon_k$.

The collision integral for 4-particle
interactions in $\lambda \phi^{4}$-theory reduces to
\begin{equation}
I_k^{(4)} = \frac{9\lambda^2}{32\pi^3\, \omega_k\, k}
\int_{0}^\infty d\varepsilon^{~}_2 
\int_{0}^\infty d\varepsilon^{~}_3 \,
D\,  F(n) \,   \, ,
\label{I4.0}
\end{equation}
where $D \equiv \text{min} [k,p_1,p_2,p_3]$ and $\varepsilon_1 =
\varepsilon_2+\varepsilon_3-\varepsilon_k >0$ in arguments of $F(n)=
\left( n^{~}_k + n^{~}_1 \right)n^{~}_2 n^{~}_3 - n^{~}_k n^{~}_1
\left( n^{~}_2+n^{~}_3 \right)$.

Note an interesting fact. Apart of the prefactor, Eqs.~\eqref{I4.0} and
\eqref{I3.0} are functions of relativistic kinetic energy, $I_k^{(i)} \equiv
(\omega_k\, k)^{-1} f(\varepsilon_k)$.  This gives for the flux of kinetic
energy in 3 dimensions (cf. with Eq.~\eqref{Frho_def})
\begin{equation}
\begin{split}
S^{\rho}({k})&=-\int^k\frac{d^{3}p}{(2\pi)^{3}}\, 
\varepsilon^{~}_p I^{(3)}_{p}\\
&=-\frac{1}{2\pi^2}\int^{\varepsilon_k} d\varepsilon\, \varepsilon\,
f(\varepsilon) \; ,
\end{split}
\label{kin-en-turb-flux}
\end{equation} 
where we have used $pdp = \omega d\varepsilon$. Therefore, the turbulent flux
should correspond to a particle distribution being a power law of relativistic
kinetic energy. Remarkably, we have solved for the turbulent fluxes without
the usual assumption of scale-independent dispersion law,
Eq.~\eqref{omega-scal}. In fact, the dispersion law was that of relativistic
field theory, $\omega_k^2 = k^2 + M^2_{\rm eff}$. We do observe a single power
law for the particle distributions as functions of kinetic energy in our
lattice integration, even in situations when $M^2_{\rm eff}$ is large in the
inertial range, see Fig.~\ref{fig:spectra}, the upper panel.

\subsection{Kinetic equation for zero-mode} 

The kinetic equation for the wave occupation numbers has to be supplemented by
the kinetic equation for the zero mode.
We start with the equation
\begin{equation}\label{zm_eq}
\ddot \phi_{0} + \omega_{c}^{2}\phi_{0}=-\lambda\left\langle\delta\phi^{3} 
\right\rangle
\end{equation}
and repeat the procedure of the previous subsections. As analog of 
Eq.~\eqref{eq-for-n} we obtain
\begin{equation}
\dot n_{c} = 2\,\lambda\,\mathrm{Im}
\left(\frac{a_{c}^{*}}{\sqrt{2\omega_{c}}}\left\langle\delta\phi^{3} 
\right\rangle\right) \; .
\end{equation}
Substituting \eqref{def:ak:inverted} and solving equation for higher order
correlators we get
\begin{eqnarray}
\dot n_{c}&\;=\;& -\int\frac{d^{d}k}{(2\pi)^{d}}\; I_{k}^{(3)} \; .
\label{set_of_kin_eq-zmode}
\end{eqnarray} 
This result is not surprising since 4-particle collisions conserve particle
number. In the model we consider, 3-particle interactions are derived from the
4-particle collisions with one of the particles being replaced by the
condensate. Therefore, Eq.~\eqref{set_of_kin_eq-zmode} can be interpreted as a
conservation of the total occupation number, in particles and in the
condensate, $n_{c}+{(2\pi)^{-d}}\int {d^{d}k}\,n_{k} = {\rm const}$.

\end{document}